\documentclass{emulateapj}
\usepackage{apjfonts}
\usepackage{lscape}
\input{epsf}
\slugcomment{Submitted to the Astronomical Journal}

\shorttitle{170 New Southern Stars with $\mu>0.45\arcsec yr^{-1}$}
\shortauthors{Sebastien Lepine}

\begin{document}

\title{New High Proper Motion Stars from the Digitized Sky Survey.
IV. Completion of the Southern Survey and 170 Additional Stars with
$\mu>0.45\arcsec yr^{-1}$.\altaffilmark{1}}

\author{S\'ebastien L\'epine}

\affil{Department of Astrophysics, Division of Physical Sciences,
American Museum of Natural History, Central Park West at 79th Street,
New York, NY 10024, USA}

\altaffiltext{1}{Based on data mining of the Digitized Sky Surveys,
developed and operated by the Catalogs and Surveys Branch of the Space
Telescope Science Institute, Baltimore, USA.}

\begin{abstract}
Completion of the SUPERBLINK proper motion survey in the southern
celestial hemisphere has turned up 170 new stars with proper motion
$0.45\arcsec yr^{-1}<\mu<2.0\arcsec yr^{-1}$. This fourth and final
installment completes the all-sky, data mining of the Digitized Sky 
Surveys for stars with large proper motions. The areas investigated 
in this final installment comprise 11,600 square degrees in the
declination range $-30^{\circ}<$Decl.$<0^{\circ}$ and in low
Galactic latitude areas south of Decl.$=-30^{\circ}$ which had not
been covered in earlier data releases. Astrometric and photometric
data are provided for the 170 new stars, along with finder
charts. Most of the new discoveries are found in densely populated
fields along the Milky Way, toward the Galactic bulge/center. The
list of new discoveries includes four stars with proper motion
$\mu>1.0\arcsec yr^{-1}$. The total list of high proper motion stars
recovered by SUPERBLINK in the southern sky now includes 2228 stars
with proper motions $0.45\arcsec yr^{-1}<\mu<2.0\arcsec
yr^{-1}$. 
\end{abstract}


\section{Introduction}

The search for stars with large proper motions has been a key research
area for stellar astrophysics in the past century. Large-scale surveys
of high proper motion (HPM) stars have been instrumental in mapping out 
the solar neighborhood, determining the kinematics of local stellar
populations, and evaluating the stellar contents of the
Galaxy. Classes of low-luminosity field stars, such as red dwarfs,
white dwarfs, and cool subdwarfs, have been first identified from
catalogs of HPM objects.

The bulk of all known stars with large proper motions was discovered
and cataloged in the course of two massive, dedicated proper motion
surveys carried out in the 1960s and 1970s. The first one is the
Lowell Proper Motion Survey, performed using the 13 inch photographic
telescope of the Lowell Observatory, using 1930s plates as a first
epoch. Stars  with proper motions $>0.25\arcsec$ yr$^{-1}$ identified
in the survey are listed in a catalog of 8,991 objects in the northern
sky \citep{G71}, and a catalog of 2758 objects in the southern sky
\citep{G78}. The second large proper motion survey is the one carried
out by W. J. Luyten, which used the 1950s National Geographic Palomar
Sky Survey as a first epoch, and a set of duplicate plates which
Luyten obtained at Palomar in the 1960s. Stars with proper motions
larger than 0.18$\arcsec$ yr$^{-1}$ were published in the ``New Luyten
Two Tenths (NLTT) catalog'' \citep{L79a}, with compiles 58,845
individual objects. The ``Luyten half-second (LHS) catalog'',
essentially a subset of the NLTT, listed stars with proper motions
$\mu>0.50\arcsec$ yr$^{-1}$, with 4470 entries \citep{L79b}. But for a few
exceptions, all the stars from the Lowell Proper Motion Survey were
also listed in the Luyten catalogs. Assembled before the generalized
use of computers, with records written by hand and typewritten, the
NLTT and LHS catalogs had recorded positional accuracies of several
arcseconds \citep{GS03} and also contained a small but significant
number of typos \citep{LSR02}. Several efforts have been made to
revise the positions of the NLTT and LHS stars, and refine the
accuracy of their astrometric and photometric measurements
\citep{BSN02,SG03} often to find some stars off their recorded
position by 1 arcmin or more.

The main limitation of the Luyten catalogs, however, was their obvious
incompleteness in specific areas of the sky \citep{D86}. The
NLTT/LHS catalogs were notably incomplete in low Galactic latitude
areas, where the high stellar density the identification of HPM stars
very challenging using the traditional blink-comparator method. The
Luyten survey was also notably incomplete in areas south of
Decl.=-30$^{\circ}$, which could not be imaged from Palomar
Observatory, from where Luyten obtained his second epoch plates. The
known incompleteness of the Luyten catalogs at southern declinations
has motivated most subsequent investigators to concentrate on the
southern sky. While the LHS/NLTT catalogs were largely complete on the
bright end ($V<15$) the potential for spectacular discoveries at the
faint end, such as very low mass stars and brown dwarfs in the
immediate proximity of the Sun, or high-velocity halo white dwarfs,
was also a strong motivation for performing more systematic and
sensitive surveys of HPM stars over the entire sky.

In the past 10 years, a succession of proper motion surveys, covering
a patchwork of southern sky areas, has been regularly supplying lists
of new HPM stars, slowly increasing the completeness of the all-sky
census. First, a survey of 131 scattered areas covering about 3200
square degrees of the southern sky was conducted by \citet{WT97} and
\citet{WC01}. The new survey identified 1642 new proper motion stars,
including 50 with $\mu>0.5\arcsec$ yr$^{-1}$. The survey was based on
a visual comparison of pairs of photographic plates on a Zeiss-Jena
blink comparator. 

A smaller survey, known as the Cal\'an-ESO (CE) Proper Motion Survey
\citep{RWRG01}, examined 325 square degrees of the southern sky in
great detail. The survey recovered 542 objects, including 385 new HPM
stars \footnote{The text of the \citet{RWRG01} paper mentions 154
stars previously listed in the Luyten catalogs; however their Table
2 clearly lists 157 stars with NLTT/LHS identifiers, from which we
obtain the number of 385 new high proper motion stars.}, of which 14
have $\mu>0.5\arcsec$ yr$^{-1}$. The survey was also based on direct
visual comparison of plate pairs using a blink machine. The survey is
famous for its discovery of Kelu-1, the first L dwarf found in
isolation \citep{RLA97}.

The Automatic Plate Measuring (APM) proper motion (APMPM) survey of
the southern sky was initiated by \citep{SIIJM00}, focusing on the
identification of stars with $0.3\arcsec$ yr$^{-1}<\mu<1.0\arcsec$
yr$^{-1}$. This survey was based on a re-analysis of scans of UK
Schmidt plates with the Cambridge APM machine. Results have however
not been published consistently, and some stars from the survey (APMPM
stars) have appeared in the literature only after follow-up
spectroscopic observations \citep{RBSO02}.

The Liverpool-Edinburgh HPM survey (LEHPM) covered over
3000 square degrees of sky near the south Galactic cap. The survey
was based on an analysis of star positions, measured with SuperCOSMOS,
in 31 fields from the R-band (IIIaF) ESO and UK Schmidt Plates. The
survey yielded 6206 detections of stars with proper motion
$\mu>0.2\arcsec$ yr$^{-1}$ \citep{PJH03}. About half of the detections
did not correspond to stars previously reported in the literature,
including a possible 92 new stars with proper motions $\mu>0.5\arcsec$
yr$^{-1}$. Some of the new objects reported in the LEHPM are however
suspected to be spurious \citep{Sub05a}.

The SuperCOSMOS-RECONS (SCR) proper motion survey was also based on
data mining of the SuperCOSMOS Sky Survey, and extended over most of
the southern sky, only excluding some areas at low Galactic
latitudes. The survey was especially productive, leading to the
discovery of 307 new stars with proper motion $\mu>0.4\arcsec$
yr$^{-1}$, including 150 with $\mu>0.5\arcsec$ yr$^{-1}$
\citep{Sub05a,Sub05b}.

Finally, the Southern Infrared Proper Motion Survey (SIPS) has been
probing the entire southern sky based on a comparison of SuperCOSMOS
Sky Survey with the Two Micron All-Sky Survey (2MASS) All-Sky Catalog,
used as a second epoch \citep{DHC05}. Thus far, 144 HPM stars have
been reported, all with $\mu>0.5\arcsec$ yr$^{-1}$, including 70 new
discoveries. Because it uses infrared data, the survey is most
sensitive to cool, very low mass stars, and brown dwarfs.

In the past seven years, we have been conducting our own survey for
stars with HPMs, with the aim of covering the entire sky, and
at the highest possible completeness rate. The SUPERBLINK proper
motion survey is based on a re-analysis of the Digitized Sky Surveys
(DSSs) using an image-differencing technique. The survey was first conducted
over the entire northern sky, before being extended to the southern
hemisphere. At northern declinations, we discovered 198 new stars with
proper motions $\mu>0.5\arcsec$ yr$^{-1}$ \citep{LSR02,LSR03}, the
largest such finding since the catalogs of Luyten. With the
publication of the LSPM-north catalog of stars with proper motions
$\mu>0.15\arcsec$ yr$^{-1}$ \citep{LS05}, the census of the northern
sky is now considered to be $99\%$ complete at high Galactic latitude
($\|b\|>15^{\circ}$) and  $90\%$ complete at low Galactic latitude
($\|b\|<15^{\circ}$), for HPM stars down to a visual magnitude of
19. In the northern sky, the LSPM-north catalog effectively
supersedes the Luyten catalogs, and should now serve as the primary
database for stars with large proper motions.

The SUPERBLINK survey now extends over most of the southern sky as
well. In the third paper of this series, we reported the discovery of
182 new stars with proper motions  $\mu>0.45\arcsec$ yr$^{-1}$, all
discovered south of  Decl.=-30$^{\circ}$. In this fourth and final
paper of the series, we report the discovery of an additional 170 new
stars with proper motions in the range $0.45\arcsec$
yr$^{-1}<\mu<2.00\arcsec$ yr$^{-1}$, identified in the final survey
area covering the remaining $\approx12,000$ square degrees of the
southern sky. This fourth list of objects completes our all-sky search
for stars with very high proper motions in the DSSs.

\begin{figure*}
\plotone{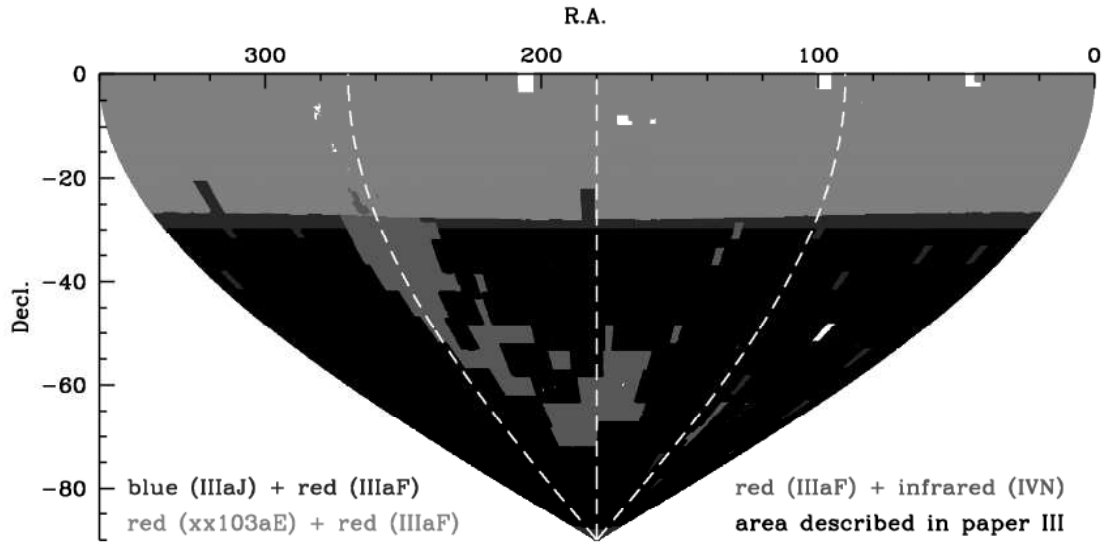}
\caption{Areas of the southern sky searched by the SUPERBLINK software
  for stars with large proper motions. Regions shaded in black
  indicate areas that were analyzed for Paper III; areas analyzed for
  this paper are shaded in color. Areas in green are regions for which we
  used DSS scans of red plates (103aE, IIIaF) for both
  the first and second epochs. Areas in blue denote regions where
  scans from blue plates (IIIaJ) were used for the first epoch, and
  scans from red plates (IIIaF) for the second epoch. Areas in red
  are those for which scans of infrared plates (IVn) were used as
  first epoch against red plates (IIIaF) for the second epoch. The
  survey overall covers $97.2\%$ of the southern sky.}
\end{figure*}

\section{Search and Identification}

HPM stars were identified from data mining of the
DSSs using the SUPERBLINK software. The method has
been described in much detail in Papers I-III of this series
\citep{LSR02,LSR03,L05} and the accuracy and completeness of the
technique, for the northern sky survey, have been documented in
\citet{LS05}. The method is based on an image subtraction algorithm,
which emphasizes objects that have moved significantly between any two
epochs of photographic images. The software generates lists of
possible moving objects along with two-epoch finder charts. These
charts are blinked on the computer screen for quality control. All
objects with very large proper motions detected by SUPERBLINK are
thus verified by human eye, which minimizes contamination by any of
several kinds of spurious detections due to plate defects and other
artifacts.

The SUPERBLINK software works by comparing pairs of images obtained at
different epochs. The efficiency in detecting HPM stars
depends on the time difference between the two images; a
longer temporal baseline will yield a higher accuracy on proper motion
measurements, and an increased sensitivity to stars with smaller
proper motions. The method also works best for fields of similar
appearance and quality (dynamical range, magnitude limit,
seeing). Ideally, the fields should have been imaged in similar
passbands and with comparable exposure times. This need not strictly
be the case, however, since SUPERBLINK includes algorithms which
adjust intensity levels and use convolutions to match the quality of
the two images as best as possible. 

The DSSs cover every part of the sky in at
least two different epochs and in various passbands (blue, red,
infrared). In the northern sky, the DSSs have full coverage at two
different epochs with sets of red plates, and with a temporal baseline
$\sim40$ yr. These are the data we have used in the north. In the
southern sky, however, the DSSs do not provide full coverage at
two epochs {\em and} in the same passband. To get full coverage at two
epochs, one has to use pairs of plates imaged in different passbands.

For about half of the southern sky, in the declination range 
$-30^{\circ}<$decl.$<0^{\circ}$, scans from the first Palomar
Observatory Sky Survey (POSS-I) are available in both blue (103aO) and
red (103aE) passbands, all imaged between 1950 and 1958. The red POSS-I
scans have been used as a first epoch against scans from the red
(IIIaF) Second Epoch Southern and Equatorial (SES) Red surveys, which cover 
the same area and were imaged between 1984 and 1998. The temporal 
baseline between the first and second epochs varied between 
27 and 47 years. The area covered by those plate pairs is shown in
Figure 1, shaded in green.

To complete the survey down to Decl.=$-30^{\circ}$, we used the same
strategy as described in Paper III (for the sky south of
$-30^{\circ}$): we used scans from the SERC-J survey as the first
epoch. These scans consist in blue (IIIaJ) plates obtained between
1974 and 1986. The SERC-J scans were matched against red (IIIaF) SES
survey scans, which consist in plates imaged between 1987 and
1998. The SES plates yield temporal baselines between 3 and 21 years,
with a median of 15 years. As explained in Paper III, the software has
the capability of matching the counterparts of moving stars from first
and second epoch images of different colors. The counterparts need not
have exactly the same plate magnitude in both images, though a large
difference due to e.g. an extreme color, will make the detection more
difficult. Figure 1 displays (shaded in blue) the area of the sky
where such blue/red plate pairs had to be used. Areas analyzed in
Paper III, using the same strategy, are shaded in black. Thirteen
additional areas south of Decl.=$30^{\circ}$ which had not been
analyzed in Paper III have been added here (also shaded in blue).

Finally, there remained areas for which no suitable first epoch images
could be found from either red or blue photographic plates. These
included a large swath of the sky at low Galactic latitude, especially
pointings toward the Galactic bulge, and also plates covering the 
Large Magellanic Cloud (LMC). For those areas, we had to use as first epoch 
scans from the SERC-I survey, which are photographic infrared plates 
obtained with the IVn emulsion and imaged between 1978 and 1988. 
These were matched against scans from the red (IIIaF) SES survey, from
plates shot between 1989 and 1996. The temporal baseline in these
red/infrared plate pairs was between 3 and 17 years, with a 12 year
median.

The southern sky was divided into 615,900 areas, each one a square
subfield $17\arcmin\times17\arcmin$ in extent. The subfields were
spaced out on a $15\arcmin\times15\arcmin$ grid, allowing for
significant overlap between neighboring subfields. SUPERBLINK
successfully processed 598,712 subfields, or 97.2\% of the total
survey area. Some 17,188 subfields were left unprocessed by
SUPERBLINK, after the software failed to properly align the
corresponding first and second epoch images. This typically occurs in
subfields containing very bright ($V<5$) stars, which saturate the
plate over an extended patch of sky, over which no reference
sources can be used for aligning the first and second epoch
images. Some extremely dense fields near the Galactic center and in
the Magellanic clouds likewise could not be processed because of
extensive plate saturation. Additional pairs of first- and second-epoch
images could not also be properly superposed by the software since it
could not match the first- and second-epoch subfields because of
significant differences between them, either due to large
differences in the seeing or significant differences in the plate
depth.

\begin{figure*}
\plotone{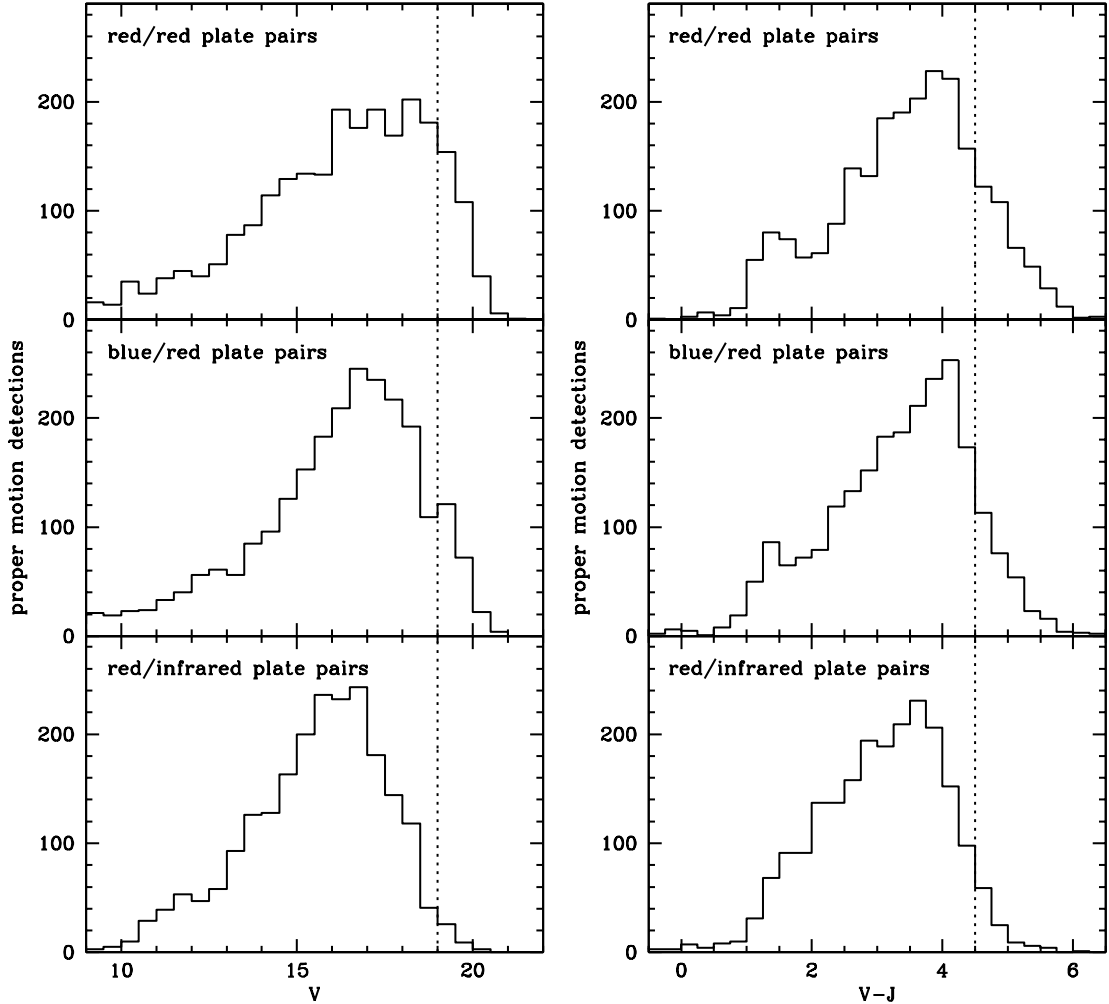}
\caption{Distribution of $V$ magnitudes and $V-J$ colors for three
  subsamples of 2500 stars identified by SUPERBLINK in each of the
  three main survey areas: regions where scans of red (IIIaF) plates
  were used for both the first and the second epoch images (red/red),
  regions where scans of blue (IIIaJ) plates were used as a first epoch
  and scans of red plates as a second epoch (blue/red), and regions
  where scans of red plates were used as a first epoch and scans of
  infrared plates (IVn) were used as a second epoch (red/infrared). The
  survey is shown to be shallower in the red/infrared areas, reaching
  only to $V=18$, while it is 2 mag deeper in the red/red
  pairs. The survey also appears to be less efficient in detecting very
  red objects ($V-J>4.5$) when using the blue/red and red/infrared
  plates.}
\end{figure*}

HPM stars were identified in all three sets of plate
pairs (red/red, blue/red, red/infrared) in roughly equivalent numbers
of detection per unit area. We did, however, observe differences in
the magnitude and color distribution of the HPM stars found in the
three survey areas. Figure 2 plots the distribution in $V$ magnitude
(left panels) and $V-J$ color (right panels) for subsamples of 2500
stars identified by SUPERBLINK in red/red, blue/red, and red/infrared
plate pairs. Our general method for estimating $V$ and $V-J$ is detailed
in \citet{LS05}. Stars detected in the red/red pairs show a sharp drop
in magnitude around $V=20$, which is consistent with the limiting
magnitude from the northern sky survey, which was also done using
red/red pairs \citep{LS05}. The $V$-magnitude distribution of stars from
the red/infrared pairs, on the other hand, shows a sharp drop at $V=18$,
suggesting that our proper motion survey is about 2 mag
shallower in the red/infrared areas. It is unclear how much of this is
due to actual limitations from the use of the different passband
red/infrared pairs, or how much could be due to the fact that most of
the red/infrared pairs covered fields at low Galactic latitude, where
the high density of stars makes the identification of faint moving
objects more difficult. Assuming all three surveys to be effectively
complete to $V=16.0$, the survey with the red/infrared pairs
is at most 80\% complete down to $V=19.0$, lacking a significant
number of the fainter ($16<V<19$) stars. The blue/red pairs survey
is at most 95\% complete down to $V=19.0$, compared with the red/red
survey.

The distributions of $V-J$ colors for stars found in the three survey
areas are relatively similar over most of the $V-J$ range, which shows
no major problems in selecting stars with moderate color terms in the
blue/red and red/infrared plates. However, the number of HPM stars
with very red colors ($V-J>4.5$) varies significantly in the three
sets. Fewer HPM stars with such a red color are detected in the blue/red
pairs (11\% of stars detected have $V-J>4.5$) than in the red/red
pairs (15\%), and even fewer are found in the red/infrared pairs
(4\%). This suggests that stars with extreme colors were not
consistently identified by the code in the blue/red and red/infrared
pairs, most probably because of significant differences in the stars'
magnitudes between the first- and second-epoch plates, due to
different imaging passbands.

\section{Recovery of known HPM stars}

In all areas of the southern sky analyzed with SUPERBLINK, we
identified a total of 2040 stars with proper motions $0.45''$
yr$^{-1}<\mu<2.00''$ yr$^{-1}$. This number excludes all the bright
stars identified by SUPERBLINK which were found listed in the TYCHO-2
catalog, which we use as a fiducial for the bright magnitude limit of
our survey. The number also includes 138 individual stars identified as
members of common proper motion pairs (69 systems). 

\begin{figure*}
\plotone{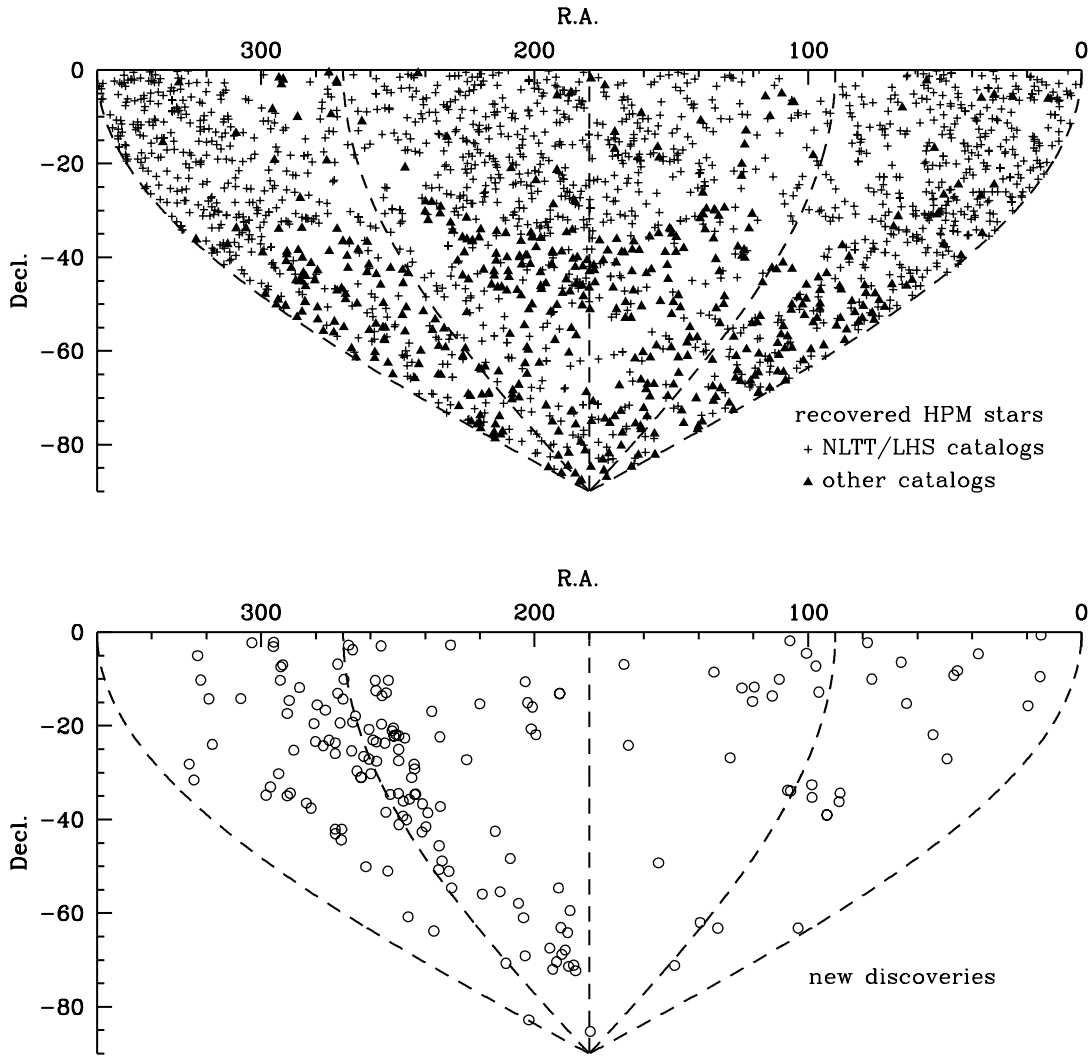}
\caption{Distribution of HPM stars identified by
  SUPERBLINK in the southern sky. Top: stars previously reported in the
  literature, including stars from Papers I-III of this
  series. Crosses mark the stars known at the time of Luyten's NLTT
  and LHS catalogs and triangles mark stars discovered in subsequent
  surveys. Bottom: new HPM stars listed in this
  paper. Many of the new discoveries are in low Galactic latitudes and
  in the general direction of the Galactic center, in fields with
  high stellar densities.}
\end{figure*}

The complete list of non-TYCHO stars was cross-correlated against all
the historical catalogs discussed in \S1. A significant fraction of
the stars detected by SUPERBLINK was found listed in one of
the either LHS or NLTT catalogs (1395 stars); of these, 234 are also listed
in the Lowell southern catalog \citep{G78}. Additionally, we found a
match to the star G267-073, which is listed in the Giclas catalog but
not in the LHS/NLTT. Among the stars that could not be matched against
the older proper motion catalogs, we found 55 matches to stars
discovered in the proper motion survey of \citet{WT97} and
\citet{WC01}. We found another 14 stars which were discovered in
the Cal\'an-ESO Proper Motion Survey \citep{RWRG01}. Some 32 stars
were first identified in the APMPM survey of
\citet{SIIJM00}, while 57 stars were first listed in the LEHPM proper
motion catalog \citep{PJH03}. Yet another 180 stars were discovered by
our own SUPERBLINK survey but reported in earlier papers of this
series \citep{LSR03,L05}, and 117 more stars were matches to objects
first reported in the SCR survey \citep{Sub05a,Sub05b}. Finally, four
more objects were first reported in the SIPS catalog \citep{DHC05}.

Stars not found in any of the catalogs above were then searched for
possible mentions in the astronomical literature using
Simbad\footnote{http://simbad.u-strasbg.fr/Simbad}. The search
yielded 15 more matches to known nearby or HPM
stars. Six objects were matches to spectroscopically confirmed white
dwarfs. The first one is the HPM white dwarf WD J0100-645
discovered by \citet{OHDHS01} in a proper motion survey of the south
Galactic cap. The second one is the HPM star USNO-B
0867-0249298, one of two stars with proper motion $\mu>1.0\arcsec$
yr$^{-1}$ confirmed by \citet{Lev05} in a follow-up analysis of HPM
entries in the USNO-B1.0 catalog of \citet{Metal03}. The other four
white dwarfs are WD 0202-055, WD 2346-478, and the common proper
motion white dwarf pair of WD 2226-754 and WD 2226-755, all reported
in the catalog of \citet{MS99}.

We also found matches to nine objects reported in the literature as very
cool, nearby stars. One was a match to APMPM J0413-3729, identified in a
search for nearby stars selected by Deep Near-Infrared Survey (DENIS)
photometry \citep{RBSO02}.  Another one is DENIS-P J210724.7-335733,
which was identified as a very nearby star in the DENIS survey
\citep{Petal01} based on its very red color and large measured proper
motion. Four other objects match the stars SSSPM J1148-7458, SSSPM
J1358-3938, SSSPM J1530-8146, and SSSPM J1549-3544, all identified by
\citet{SLMZ04} in a proper motion survey combining the SuperCOSMOS Sky
Survey database and positions from the 2MASS and DENIS surveys. We
also found a match to the late-type M8 dwarf 2MASSI J1534570-141848,
which was observed as part of a spectroscopic backup program by
\citet{G02}. This star was selected for follow-up observation because
of a very red infrared color $J-K_s>1.0$, indicative of a very
low mass star, so it was not initially identified as a HPM object. We
find this star to have a very large proper motion
$[\mu_{RA},\mu_{Decl.}]=[-0.931,-0.317] \arcsec$ yr$^{-1}$, and
although the star cannot formally be considered a ``new discovery,''
ours is the first report of a proper motion measurement. Likewise
we recovered the star 2MASSI 0417374-080000, also initially identified
as a candidate nearby star ($d\simeq17.4$ pc) from a color-selected
subsample of sources from the 2MASS Second Instrumental Release
\citep{CRLKL03}. We find the star to have a proper motion
$[\mu_{RA},\mu_{Decl.}]=[0.458,0.041] \arcsec$ yr$^{-1}$. Finally, the
esdM7.5 subdwarf 2MASS J12270506-0447207 was also recovered by
SUPERBLINK, with a proper motion
$[\mu_{RA},\mu_{Decl.}]=[-0.459,0.249] \arcsec$ yr$^{-1}$; this star
had been recently identified as an ultra-cool subdwarf from a
color-selected sample of 2MASS sources \citep{BCK07}.

The location on the sky of all the stars with $0.45\arcsec$
yr$^{-1}<\mu<2.00\arcsec$ yr$^{-1}$ identified by SUPERBLINK is
shown in Figure 3 (top panel). Stars which were listed in the NLTT/LHS
catalogs are marked with crosses. Stars discovered in subsequent
surveys, as described above, are plotted as triangles.

\section{New discoveries}

In the end, the list of SUPERBLINK identifications yielded 170 new
stars with $0.45\arcsec$ yr$^{-1}<\mu<2.0\arcsec$ yr$^{-1}$. Their
distribution on the sky is shown in Fig.3 (bottom panel). A
significant fraction of these new discoveries are located in low
Galactic latitude areas, particularly in the general direction of the
Galactic center/bulge (around $\alpha$=270.0 $\delta$=-30.0). This
reflects the particular ability of the SUPERBLINK software to detect
moving stars in fields of high stellar density, areas which were
typically avoided in most previous surveys. These areas of low
Galactic latitude in the southern sky were the very last regions of
the sky which remained largely unexplored for faint HPM
stars. Our survey has now filled this gap.

The complete list of 170 new HPM stars is provided in
Table 1. All stars from the SUPERBLINK survey are now assigned a
designation following the convention introduced in Paper III of this
series, which is based on the system initially proposed by
\citet{E79}. The prefix ``PM'' (for proper motion) is followed by an
alpha-numerical designation which refers to the star's
coordinates. The sequence starts with the letter ``I'' which indicates
that the coordinates are from the International Celestial Reference
System (ICRS) (all positions are based on 2MASS catalog positions of
the infra-red counterparts, which are ICRS resolved). A directional
suffix (``N,'' ``S,'' ``W,'' or ``E'') is added to distinguish between
stars with similar coordinates. Table 1 gives the right ascension
(R.A.) and decl. for every star ($\alpha$(ICRS), $\delta$(ICRS) ). The
total proper motion of the star ($\mu$) is given in arcseconds per year,
as well as the component of the proper motion in the directions of
R.A. and decl. ($\mu_{\alpha}$, $\mu_{\delta}$). The temporal baseline
($\Delta t$) between the first and second epoch detections is
tabulated, from which an estimate of the proper motion error is
calculated ($\mu_{err}$).

We searched for counterparts to all the HPM stars
in the USNO-B1.0 Catalog \citep{Metal03}, from which we extracted
optical magnitudes in the blue (B$_J$) red (R$_F$) and infrared
(I$_N$) bandpasses. Counterparts were also identified in the 2MASS
All-Sky Catalog \citep{C03}, yielding infrared $J$, $H$, and $K_s$
magnitudes for most of the stars. All the magnitudes are listed in
Table 1. Effective visual magnitudes ($V$) were then calculated from the
photographic magnitudes using the scheme of \citet{LS05}. The table
then lists optical-to-infrared ($V-J$) colors for all stars with 2MASS
counterparts.

\begin{figure}
\epsscale{2.4}
\plotone{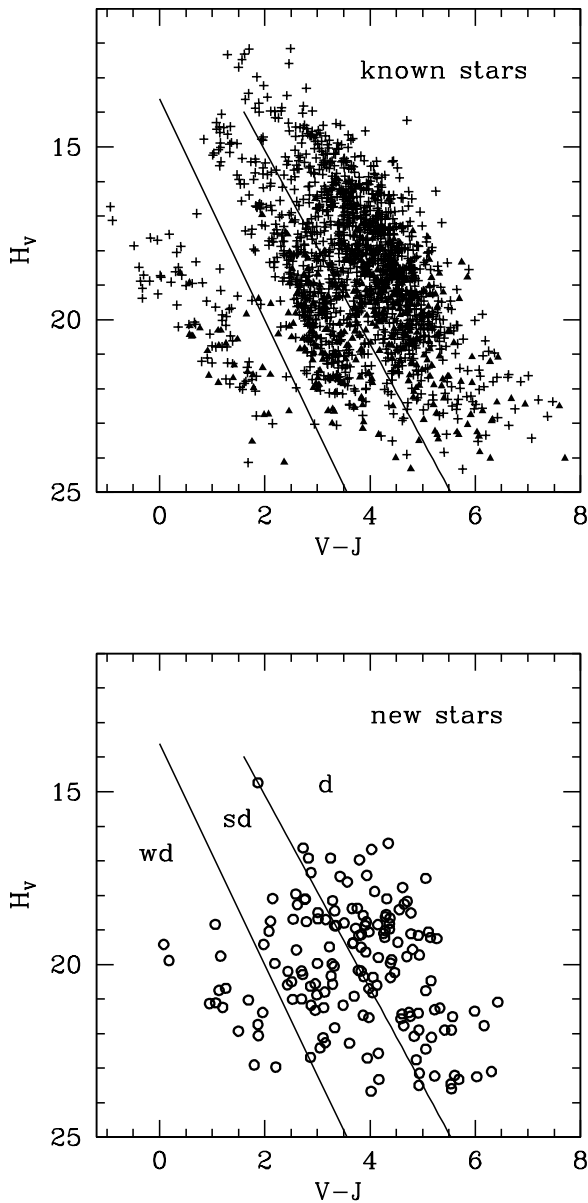}
\caption{Reduced proper motion diagrams of the stars identified by
  SUPERBLINK. Top: previously known HPM stars. Labels have the same
  meaning as in Fig.3, with crosses denoting stars from the NLTT/LHS
  catalogs and triangles stars that were discovered since the Luyten
  surveys. Bottom: new HPM stars discovered with
  SUPERBLINK. The location on the diagram is used to separate the
  stars in one of three classes: white dwarfs (wd), subdwarfs (sd),
  and dwarfs (d), which occupy distinct loci \--- most apparent in the
  top panel. The subdwarfs are typically associated with the Galactic
  halo and the dwarfs with the Galactic disk.}
\end{figure}

A reduced proper motion diagram, plotting $H_{V}={V}+5\log{\mu}+5$
against $V-J$, is displayed in Figure 4. The top panel shows the stars
from earlier catalogs which have been re-identified in our survey; the
bottom panel shows the new discoveries. Nearby stars placed in a
reduced proper motion diagram occupy one of three distinct loci. Data
from our extensive spectroscopic follow-up survey of northern stars
with large proper motions \citep{LRS03,LRS07} demonstrate that these
loci correspond to three distinct spectroscopic classes: low-mass dwarfs
(d), low-mass subdwarfs (sd), and white dwarfs (wd). The three classes
are arranged in three main ``layers'' from the upper right to the
bottom left of the diagram, respectively, and roughly match the
relative locations of the color-magnitude relationships, as one would
see them in a color-magnitude diagram. We set the approximate
boundaries between the three loci using two straight lines, shown in
Fig.4. We use these boundaries to assign each of our stars one of
three spectral classes (d/sd/wd). These are essentially predictions of
the spectral subclasses, meant to be used only as a guide for future
spectroscopic follow-up efforts. Based on our ongoing follow-up
spectroscopic survey of over 3000 stars, this predictor based on the
reduced proper motion is about 90\% accurate. The spectral class
assignments (d/sd/wd) for the 170 new stars are listed in the last
column of Table 1.

\section{Notes on individual stars}

\subsection{PM I16230-6905}

With a total proper motion $\mu=1.594\arcsec$ yr$^{-1}$, this star is
the fastest of the newly discovered objects. Its location in the
reduced proper motion diagram places it within the group of M dwarfs
from the disk population. However its $V$ magnitude and $V-J$ color yield
a photometric distance of $\sim29$ parsecs, following the
color-magnitude relationship of \citet{L05}. This would indicate a
large transverse velocity of $\sim220$ km s$^{-1}$. Such a large
velocity is more usually associated with subdwarfs from the halo. It
is very possible that the star is moderately metal poor (log Fe/H
$\approx$ -0.5) and also slightly sub-luminous; this will give it a
relatively closer photometric distance and smaller transverse
velocity. This would be enough to kinematically associate the star
with the thick disk population, which would also be consistent with
the star being moderately metal poor.

\subsection{PM I17189-4131}

This star has a total proper motion $\mu=1.160\arcsec$ yr$^{-1}$. Its
very red color ($V-J=5.51$, $J-K_s=0.99$) suggests it is a
late-type red dwarf, probably of spectral subtype M6. The photometric
distance, again following the [M$_{V}$,$V-J$] color-magnitude relationship
of \citet{L05}, places it at $d\approx18$ parsecs. Just like the
previous object, the very high PM would then indicate a
relatively large transverse velocity $v_T\approx100$
km s$^{-1}$. Spectroscopy is much needed in this case, as it is
required to determine whether the star is metal poor and a member of
the old disk or thick disk, or whether it could be a kinematically hot
member of the thin disk population.

\subsection{PM I17595-1755}

With a proper motion $\mu=1.143\arcsec$ yr$^{-1}$, this star has
a very large reduced proper motion for an object of that color
($V-J\simeq3.3$) so much so that its location in the reduced proper
motion diagram is largely consistent with the halo subdwarf
population. Spectral classification would be required to obtain a
photometric distance estimate, because color-magnitude relationships
for the cool subdwarfs are dependent on metallicity. 

\subsection{PM I19418-0208}

The fourth new star with a proper motion in excess of 1$\arcsec$
yr$^{-1}$, this one is remarkably faint ($V\simeq17.5$), and not
particularly red ($V-J\simeq2.87$). Its location in the reduced proper
motion diagram lies just within the subdwarf range, although there is
a chance that the star could be a white dwarf. This is yet another
high-priority target for spectroscopic follow-up observations.

\section{Missed objects and survey completeness}

In order to assess the completeness of our proper motion survey, we
have been compiling hits and misses for stars identified in all the
previous surveys. Here we analyze the results for stars with
proper motions in the range $0.45\arcsec$ yr$^{-1}<\mu<2.0\arcsec$
yr$^{-1}$ and excluding the stars which are listed in the TYCHO-2
catalog (i.e. the very bright stars).

To determine the recovery rate and efficiency of the SUPERBLINK
software for the southern sky areas, we have compared the complete
list of SUPERBLINK detections with the following catalogs and lists of
HPM objects:
\begin{enumerate}
\item The Lowell southern proper motion catalog of \citet{G78}
\item The NLTT catalog of \citet{L79a}.
\item The lists of HPM stars identified by \citet{WT97}
  and \citet{WC01}, which we also refer to as the WT/WC survey
\item The Cal\`an-ESO Proper Motion Survey \citep{RWRG01}.
\item The Liverpool-Edinburgh HPM Survey \citep{PJH03},
  which we also refer to as the LEHPM survey.
\item The APMPM survey of \citet{SIIJM00}.
\item The SuperCOSMOS-RECONS (SCR) proper motion survey
  \citep{Sub05a,Sub05b}.
\item The Southern Infrared Proper Motion Survey or SIPS \citep{DHC05}.
\end{enumerate}
We have carefully cross-correlated all southern sky objects from the
catalogs above to identify overlaps and eliminate redundant
objects. Many stars were indeed found to be listed in more than one of
the catalogs above. As a rule, of course, we only considered stars
from these surveys which are not listed in the TYCHO-2 catalog, and
with proper motions within the limits of our current data
release. Together, the eight catalogs provided a total of 1962
non-TYCHO stars with proper motion $0.45\arcsec$
yr$^{-1}<\mu<2.0\arcsec$ yr$^{-1}$.

A cross-correlation with the SUPERBLINK list has yielded matches
to 1730 objects; 232 stars from the catalogs above were not
identified by SUPERBLINK. To verify the reality of the missing stars,
we have examined DSS scans centered on the quoted
positions of each of the 232 objects. All fields were 10$\arcmin$ on
the side. We were able to confirm the HPM status of 210
stars, but failed to identify 22 objects (21 of which were listed only
once in the NLTT catalog, and the other one mentioned only once in the
LEHPM survey); we must assume these stars to be either spurious or to
have such large errors on their reported positions ($>5\arcmin$) that
the stars are effectively ``lost.'' In the end, we find that our
SUPERBLINK survey failed to recover 210 genuine HPM
stars out of 1940 genuine objects, suggesting an overall recovery
rate of 89.2\%. 

We compile in Table 2 the number of hits and misses from each of the
older surveys, and calculate the effective rate of recovery by
SUPERBLINK. A disproportionate number of misses are from the SIPS
survey, from which SUPERBLINK recovered only 20 of the 68 stars. The
SIPS survey was based on a comparison of infrared images, and was
thus particularly sensitive to extremely red objects, particularly
brown dwarfs; objects like these often do not show up on the optical
images of the DSSs. Because SUPERBLINK always used at
least one of the blue and red photographic plates, these objects could
not possibly have been detected. If we exclude misses from the SIPS
catalog, then we find an overall recovery rate of 91.4\% for SUPERBLINK. 

\begin{figure}
\epsscale{2.5}
\plotone{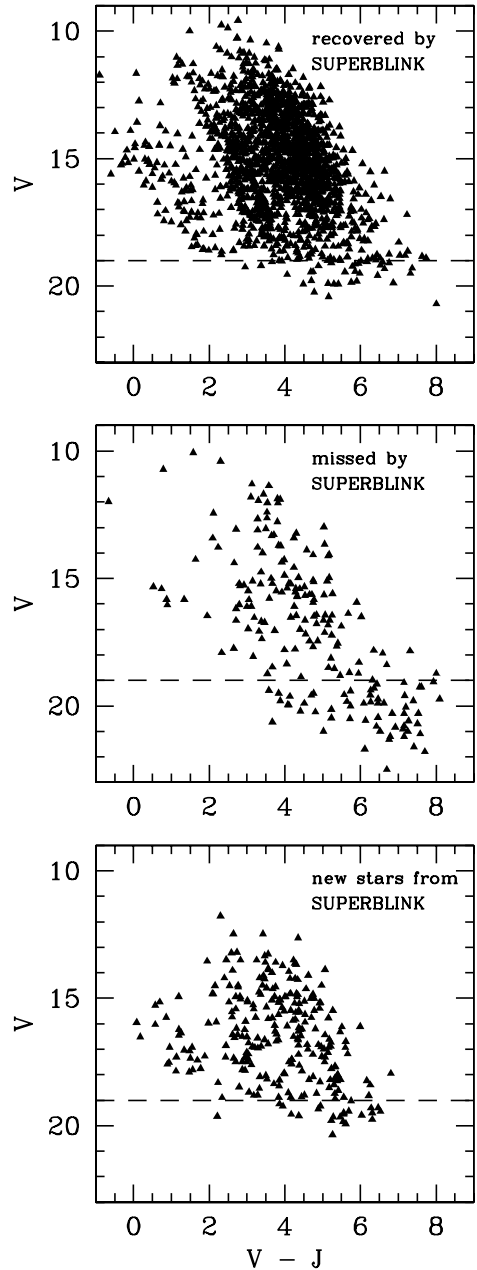}
\caption{Distribution of apparent visual magnitudes ($V$)
  and optical-to-infrared color ($V-J$) of HPM stars
  ($0.45\arcsec$ yr$^{-1}<\mu<2.0\arcsec$ yr$^{-1}$ identified in
  the entire southern celestial hemisphere (excluding bright TYCHO-2
  catalog objects). Top panel: known stars re-identified by
  SUPERBLINK (``hits''). Center: known stars that were not recovered
  by SUPERBLINK (``misses''). Bottom: new stars identified with
  SUPERBLINK, and listed in Table 1 of this paper. The recovery rate
  by SUPERBLINK of previously known stars is larger than 90\% for
  visual magnitudes $V<19$ (dashed line).}
\end{figure}

Figure 5 shows the color-magnitude distribution of the stars recovered
and missed by SUPERBLINK. The top panel shows the recovered objects
and the middle panels the stars missed by the code. SUPERBLINK has a
recovery rate exceeding 90\% for all stars with $V$ magnitude brighter
than 19.0. We take this to be the magnitude limit of our survey, and
denote it with a dashed line in Fig.5. Stars with $V<19$ that were
missed by SUPERBLINK have a distribution of magnitudes and colors
which matches the distribution from recovered objects. This indicates
that most of the misses are not related to any magnitude or color
selection effects. Most of the misses are located at high Galactic
latitudes ($|b|>20$), but this is probably just because earlier
surveys did not have a good coverage of the low Galactic
latitudes. 

Rather, objects are probably missed when SUPERBLINK fails to properly
process specific pairs of photographic plate scans. The software uses
$17\arcmin\times17\arcmin$ subfileds, which can be improperly processed for a
variety of reasons. One is the presence of an extremely bright
($V\lesssim5$) saturated star in the field which makes it difficult to
properly align the first and second epoch images. Problems also often
arise in fields which are extremely sparse, because the software
requires a minimum number of background sources in the
$17\arcmin\times17\arcmin$ subfields to again properly align the first-
and second-epoch images. Additional problems arise when images from
each of the two epochs are of very different qualities, particularly if
there are significant differences in the seeing or magnitude
depth. All these possible problems interfere with the detection of
HPM stars in seemingly randomly distributed fields in
the southern sky, which would yield misses with no obvious trend in
color or magnitude.


\section{Conclusions}

Our hunt for stars with very large proper motions detected in the
DSSs is now coming to a close. We are confident
that the vast majority of H-burning stars with proper motions
$0.45\arcsec$ yr$^{-1}<\mu<2.0\arcsec$ yr$^{-1}$ have now been
identified over the entire sky, which completes the efforts undertaken
by W. J. Luyten over 60 years ago. The 170 new stars presented in this
paper most likely represent the very last, large contingent of new stars
with very large proper motions. We find that the SUPERBLINK survey was
particularly useful in locating missing HPM stars in
dense fields at low Galactic latitudes, although it is likely that a
few stars remain to be identified in the densest Milky Way fields.

The reduced proper motion diagram shows that the new stars are a mix
of nearby disk dwarfs, halo subdwarfs, and white dwarfs. We
tentatively assign each of the new stars to one of the three main
classes. It is however clear from the few examples in \S5 that
spectroscopic follow-up is required to ascertain the status of several
of the stars. Likewise, many of the previously known HPM stars in the
southern sky still have no formal spectral classification, and should
be considered high priority targets.

Lists of southern stars with HPMs are unfortunately
still scattered throughout the astronomical literature, in no less than
eight separate publications. The upcoming LSPM-south catalog (S. L\'epine
{\it et al.} 2008, in preparation) will remedy this problem by combining
all the stars in a single catalog. We now have all-sky lists of
SUPERBLINK detections down to proper motions of $0.15\arcsec$
yr$^{-1}$. What needs to be done now is the independent validation of
the HPM stars reported in other surveys but missed by
SUPERBLINK.

In any case, the census of stars with very large proper motions
$\mu>0.45\arcsec$ yr$^{-1}$ is now largely complete down to magnitude
$V=19.0$. The census of objects with very large proper motion is
however still incomplete at fainter magnitudes. Objects that await
discovery most likely include a number of old white dwarfs and
ultra-cool dwarfs and subdwarfs, all too faint to be detected in the
DSSs. It is also likely that brighter HPM stars are
still hiding in very dense fields, or at short angular separations from
very bright stars, since their images would have been engulfed within
the photographically saturated patches of their very bright
line-of-sight companions. However, we suspected that the bulk of the
still-missing HPM objects most probably consists of all
types of brown dwarfs and other extremely red and low-luminosity
objects. Their discovery will require deep, multi-epoch surveys,
preferably conducted at infrared wavelengths.


\acknowledgments

{\bf Acknowledgments}

This research program was supported by the National Science Foundation
through grant AST-0607757, held at the American Museum of Natural
History. S.L. also gratefully acknowledges support from Hilary Lipsitz.

This work has been made possible through the use of the Digitized Sky
Surveys. The Digitized Sky Surveys were produced at the Space
Telescope Science Institute under U.S. Government grant NAG
W-2166. The images of these surveys are based on photographic data
obtained using the Oschin Schmidt Telescope on Palomar Mountain and
the UK Schmidt Telescope. The plates were processed into the present
compressed digital form with the permission of these institutions. The
National Geographic Society - Palomar Observatory Sky Atlas (POSS-I)
was made by the California Institute of Technology with grants from
the National Geographic Society. The Second Palomar Observatory Sky
Survey (POSS-II) was made by the California Institute of Technology
with funds from the National Science Foundation, the National
Geographic Society, the Sloan Foundation, the Samuel Oschin
Foundation, and the Eastman Kodak Corporation. The Oschin Schmidt
Telescope is operated by the California Institute of Technology and
Palomar Observatory. The UK Schmidt Telescope was operated by the
Royal Observatory Edinburgh, with funding from the UK Science and
Engineering Research Council (later the UK Particle Physics and
Astronomy Research Council), until 1988 June, and thereafter by the
Anglo-Australian Observatory. The blue plates of the southern Sky
Atlas and its Equatorial Extension (together known as the SERC-J), as
well as the Equatorial Red (ER), and the Second Epoch red Survey
(SES) were all taken with the UK Schmidt. 

This publication makes use of data products from the Two Micron All
Sky Survey, which is a joint project of the University of
Massachusetts and the Infrared Processing and Analysis
Center/California Institute of Technology, funded by the National
Aeronautics and Space Administration and the National Science
Foundation.

The data mining required for this work has been made possible with the
use of the SIMBAD astronomical database and VIZIER astronomical
catalogs service, both maintained and operated by the Centre de
Donn\'ees Astronomiques de Strasbourg\footnote{http://cdsweb.u-strasbg.fr/}.

This research used the facilities of the Canadian Astronomy Data 
Centre\footnote{http://www4.cadc-ccda.hia-iha.nrc-cnrc.gc.ca/cadc/} 
operated by the National Research Council of Canada with the support 
of the Canadian Space Agency.


\clearpage
\begin{landscape}
\begin{deluxetable}{lrrrrrrrrrrrrrrrr}
\tabletypesize{\scriptsize}
\tablecolumns{17}
\tablewidth{0pt}
\tablecaption{New High proper motion stars.}
\tablehead{
\colhead{Star} & 
\colhead{$\alpha$(ICRS)} &
\colhead{$\delta$(ICRS)} &
\colhead{$\mu$} & 
\colhead{$\mu_{\alpha}$} & 
\colhead{$\mu_{\delta}$} & 
\colhead{$\Delta t$\tablenotemark{a}} & 
\colhead{$\mu_{err}$\tablenotemark{b}} & 
\colhead{$B_J$\tablenotemark{c}} & 
\colhead{$R_F$} &
\colhead{$I_N$} & 
\colhead{$J$\tablenotemark{d}} &
\colhead{$K$}&
\colhead{$H$}&
\colhead{$V$}&
\colhead{$V-J$}&
\colhead{class}\\
\colhead{} & 
\colhead{} &
\colhead{} &
\colhead{($\arcsec$ yr$^{-1}$)} & 
\colhead{($\arcsec$ yr$^{-1}$)} & 
\colhead{($\arcsec$ yr$^{-1}$)} & 
\colhead{(yr)} & 
\colhead{($\arcsec$ yr$^{-1}$)} & 
\colhead{} & 
\colhead{} &
\colhead{} & 
\colhead{} &
\colhead{} &
\colhead{} &
\colhead{} &
\colhead{} &
\colhead{}
}
\startdata 
PM I00432-6310 &  10.814372& -63.178169&  0.699&  0.264& -0.647&  10.0& 0.040& 99.8& 14.1& 12.9& 12.74& 12.22& 12.06&  14.82&  2.08& sd\\
PM I00513-0930 &  12.846331&  -9.501763&  0.571& -0.096& -0.563&  35.1& 0.011& 20.1& 18.9& 17.1& 15.38& 14.92& 14.74&  19.55&  4.17& sd\\
PM I00532-1542 &  13.324378& -15.710962&  0.527&  0.517& -0.101&  37.7& 0.011& 18.2& 15.5& 14.4& 13.67& 13.10& 12.90&  16.96&  3.29& sd\\
PM I00591-0038 &  14.794205&  -0.647302&  0.461&  0.084& -0.453&  31.9& 0.013& 19.4& 18.1& 17.4& 15.69& 15.59& 15.19&  18.80&  3.11& sd\\
PM I02124-2703 &  33.124096& -27.060802&  0.697& -0.043& -0.696&  18.9& 0.021& 20.0& 19.0& 99.8& 99.99& 99.99& 99.99&  19.54& 99.99&  ?\\
PM I02289-0439 &  37.230431&  -4.662395&  0.546&  0.054& -0.543&  40.7& 0.010& 20.4& 17.1& 15.4& 14.72& 14.23& 13.93&  18.88&  4.16& sd\\
PM I02560-0813 &  44.009895&  -8.220679&  0.453& -0.044& -0.451&  37.9& 0.011& 20.8& 18.9& 16.2& 14.32& 13.75& 13.39&  19.93&  5.61&  d\\
PM I02583-2153 &  44.582863& -21.890055&  0.539&  0.508&  0.180&  43.7& 0.009& 20.1& 19.0& 15.1& 13.56& 13.06& 12.71&  19.59&  6.03&  d\\
PM I02597-0913 &  44.949638&  -9.219873&  0.604&  0.296& -0.527&  37.8& 0.011& 16.8& 13.3& 12.3& 10.93& 10.42& 10.14&  15.19&  4.26&  d\\
PM I03588-1515 &  59.710186& -15.252027&  0.484&  0.375& -0.306&  34.7& 0.012& 21.0& 18.4& 15.9& 14.57& 14.24& 13.84&  19.80&  5.23&  d\\
PM I04210-0625 &  65.254318&  -6.429371&  0.532&  0.019& -0.532&  35.9& 0.011& 21.9& 17.4& 16.1& 14.29& 13.81& 13.45&  19.83&  5.54&  d\\
PM I04269-3612 &  66.744637& -36.212078&  0.471&  0.469& -0.038&  15.0& 0.027& 99.8& 15.1& 12.6& 11.39& 10.86& 10.54&  16.20&  4.81&  d\\
PM I04354-3422 &  68.871231& -34.368519&  0.495& -0.164& -0.467&  15.0& 0.027& 99.8& 12.4& 10.7&  9.70&  9.15&  8.88&  13.50&  3.80&  d\\
PM I05002-1000 &  75.056885& -10.016049&  0.901&  0.491& -0.755&  39.9& 0.010& 16.8& 13.8& 13.2& 12.72& 12.28& 12.08&  15.42&  2.70& sd\\
PM I05030-6311 &  75.765450& -63.185085&  0.528&  0.305&  0.431&  11.9& 0.034& 19.4& 16.2& 15.1& 13.95& 13.44& 13.25&  17.93&  3.98& sd\\
PM I05127-0214 &  78.186485&  -2.237835&  0.482&  0.332& -0.350&  40.0& 0.010& 20.4& 18.1& 16.3& 14.46& 13.98& 13.71&  19.34&  4.88&  d\\
PM I05208-3518 &  80.218033& -35.304283&  0.466&  0.351&  0.306&  15.7& 0.025& 15.4& 12.6& 11.8& 10.68& 10.10&  9.83&  14.11&  3.43&  d\\
PM I05334-7110 &  83.369072& -71.172531&  0.489&  0.478& -0.104&  14.0& 0.029& 16.8& 16.2& 14.6& 13.52& 12.99& 12.76&  16.52&  3.00& sd\\
PM I05338-3234 &  83.465843& -32.573948&  0.481&  0.122& -0.465&  15.0& 0.027& 20.1& 18.9& 18.6& 99.99& 99.99& 99.99&  19.55& 99.99&  ?\\
PM I06052-3354 &  91.304489& -33.904339&  0.454& -0.189&  0.413&  15.9& 0.025& 19.4& 17.7& 14.7& 13.20& 12.61& 12.25&  18.62&  5.42&  d\\
PM I06107-3346 &  92.687683& -33.772923&  0.766&  0.601&  0.475&  15.9& 0.025& 18.1& 15.9& 13.6& 11.52& 10.94& 10.66&  17.09&  5.57&  d\\
PM I06149-6158 &  93.740448& -61.982430&  0.479&  0.156&  0.453&  15.0& 0.027& 99.8& 17.0& 14.7& 13.33& 12.89& 12.53&  18.10&  4.77&  d\\
PM I06157-1247 &  93.931671& -12.789595&  0.618&  0.452& -0.421&  39.2& 0.010& 20.7& 18.2& 16.6& 14.62& 14.13& 13.68&  19.55&  4.93& sd\\
PM I06256-0715 &  96.414886&  -7.258792&  0.478& -0.106& -0.466&  41.1& 0.010& 16.7& 14.8& 13.0& 11.55& 10.99& 10.67&  15.83&  4.28&  d\\
PM I06411-0432 & 100.299698&  -4.536717&  0.906&  0.751&  0.506&  31.0& 0.013& 18.1& 16.3& 16.0& 15.39& 15.35& 15.24&  17.27&  1.88& wd\\
PM I07063-0151 & 106.585503&  -1.859489&  0.469&  0.355& -0.307&  33.8& 0.012& 20.2& 17.0& 14.9& 13.89& 13.41& 13.14&  18.73&  4.84&  d\\
PM I07177-1004 & 109.427681& -10.080668&  0.510&  0.314& -0.402&  27.0& 0.015& 19.2& 17.1& 99.8& 12.06& 11.48& 11.15&  18.23&  6.17&  d\\
PM I07245-1338 & 111.145668& -13.641150&  0.488&  0.158& -0.462&  27.2& 0.015& 18.0& 16.5& 16.8& 16.18& 16.15& 15.61&  17.31&  1.13& wd\\
PM I07530-1447 & 118.274628& -14.786615&  0.534& -0.199&  0.496&  32.2& 0.012& 17.1& 15.2& 13.2& 11.82& 11.28& 11.04&  16.23&  4.41&  d\\
PM I07534-1142 & 118.358902& -11.700098&  0.753&  0.209& -0.723&  29.0& 0.014& 18.1& 15.9& 15.2& 13.22& 12.74& 12.54&  17.09&  3.87& sd\\
PM I08089-2649 & 122.229820& -26.827749&  0.485&  0.191& -0.446&  42.1& 0.010& 16.7& 14.2& 12.7& 11.18& 10.68& 10.43&  15.55&  4.37&  d\\
PM I08117-1155 & 122.937828& -11.918666&  0.466& -0.150&  0.441&  29.1& 0.014& 16.4& 14.4& 13.6& 11.11& 10.60& 10.30&  15.48&  4.37&  d\\
PM I08555-0833 & 133.894638&  -8.562494&  0.467& -0.075& -0.461&  38.1& 0.010& 18.5& 17.5& 17.2& 16.08& 15.84& 15.28&  18.04&  1.96& wd\\
PM I09248-4915 & 141.204376& -49.258228&  0.910& -0.694&  0.588&  14.8& 0.027& 17.0& 15.8& 15.8& 15.25& 14.92& 14.80&  16.45&  1.20& wd\\
PM I10573-2411 & 164.337021& -24.184950&  0.605& -0.604& -0.032&  12.1& 0.033& 20.3& 18.4& 15.7& 13.74& 13.21& 12.86&  19.43&  5.69&  d\\
PM I11088-0655 & 167.221176&  -6.924221&  0.641& -0.596&  0.237&  30.8& 0.013& 20.1& 18.4& 17.8& 99.99& 99.99& 99.99&  19.32& 99.99&  ?\\
PM I11425-8515 & 175.638351& -85.262344&  0.473& -0.469&  0.060&  14.9& 0.027& 18.4& 18.5& 17.9& 99.99& 99.99& 99.99&  18.45& 99.99&  ?\\
PM I12445-1306S& 191.132065& -13.107369&  0.511& -0.476& -0.186&  36.1& 0.011& 16.2& 13.9& 12.5& 12.13& 11.54& 11.34&  15.14&  3.01& sd\\
PM I12445-1306N& 191.133575& -13.104939&  0.511& -0.476& -0.186&  36.1& 0.011& 15.8& 13.1& 12.1& 11.79& 11.25& 11.03&  14.56&  2.77& sd\\
PM I12552-5928 & 193.800598& -59.471436&  0.503&  0.137& -0.484&  12.8& 0.031& 19.1& 15.8& 13.9& 11.15& 10.58& 10.21&  17.58&  6.43&  d\\
PM I13058-7218 & 196.465515& -72.302284&  0.456& -0.335&  0.310&  17.8& 0.022& 17.9& 16.7& 16.6& 14.70& 14.24& 14.07&  17.21&  2.51& sd\\
PM I13108-7107 & 197.716721& -71.119659&  0.561& -0.510& -0.234&  17.8& 0.022& 15.8& 13.8& 12.0& 10.59& 10.06&  9.72&  14.88&  4.29&  d\\
PM I13126-6408 & 198.169327& -64.142029&  0.487& -0.485&  0.041&  14.0& 0.029& 99.8& 15.3& 15.1& 14.00& 13.58& 13.30&  15.98&  1.98& sd\\
PM I13174-5438 & 199.364120& -54.635632&  0.732& -0.569& -0.461&  13.2& 0.030& 17.1& 15.5& 15.8& 15.10& 14.76& 14.65&  16.36&  1.26& wd\\
PM I13240-2153 & 201.010391& -21.884447&  0.483& -0.455& -0.162&  38.9& 0.010& 20.4& 19.1& 18.7& 99.99& 99.99& 99.99&  19.80& 99.99&  ?\\
PM I13265-1558 & 201.631165& -15.979954&  0.505& -0.505& -0.011&  34.9& 0.011& 18.4& 17.9& 17.9& 99.99& 99.99& 99.99&  18.17& 99.99&  ?\\
PM I13306-2039 & 202.670593& -20.650953&  0.591&  0.316& -0.499&  39.8& 0.010& 13.6& 11.5&  9.4&  8.28&  7.65&  7.39&  12.63&  4.35&  d\\
PM I13317-6305 & 202.947296& -63.085049&  0.626& -0.609&  0.147&   6.3& 0.063& 99.8& 16.1& 13.6& 13.71& 13.17& 12.92&  17.20&  3.49& sd\\
PM I13332-6751 & 203.310654& -67.854759&  0.795& -0.743& -0.283&  12.9& 0.031& 17.7& 16.7& 16.0& 15.37& 15.20& 13.37&  17.24&  1.87& wd\\
PM I13334-1503 & 203.355637& -15.061091&  0.548& -0.457& -0.302&  28.8& 0.014& 19.4& 16.8& 14.8& 13.27& 12.84& 12.45&  18.20&  4.93&  d\\
PM I13341-7121 & 203.547928& -71.364250&  0.502& -0.376& -0.333&  12.9& 0.031& 15.9& 13.4& 11.5& 10.10&  9.51&  9.20&  14.75&  4.65&  d\\
PM I13351-1037 & 203.794006& -10.618372&  0.464& -0.160&  0.436&  36.8& 0.011& 20.6& 18.5& 18.2& 17.42& 16.11& 15.50&  19.63&  2.21& wd\\
PM I13506-6847 & 207.663071& -68.788033&  0.746& -0.595& -0.450&  12.9& 0.031& 14.6& 13.1& 12.0& 11.29& 10.69& 10.40&  13.91&  2.62& sd\\
PM I14223-7023 & 215.581116& -70.393639&  0.819& -0.726& -0.380&  13.1& 0.031& 16.5& 13.6& 11.2& 10.23&  9.67&  9.37&  15.17&  4.94&  d\\
PM I14309-6728 & 217.735794& -67.478683&  0.564& -0.436& -0.358&  10.3& 0.039& 17.5& 15.6& 14.0& 12.24& 11.69& 11.40&  16.63&  4.39&  d\\
PM I14460-1518 & 221.509415& -15.304267&  0.465& -0.463&  0.046&  39.0& 0.010& 20.5& 18.9& 15.6& 13.45& 12.88& 12.57&  19.76&  6.31&  d\\
PM I14521-7157 & 223.025345& -71.962532&  0.491& -0.465& -0.159&  13.1& 0.031& 17.6& 15.8& 13.5& 12.30& 11.82& 11.50&  16.77&  4.47&  d\\
PM I14537-4820 & 223.430466& -48.345669&  0.501& -0.487& -0.116&  13.2& 0.030& 18.6& 16.2& 16.0& 14.80& 14.15& 14.05&  17.50&  2.70& sd\\
PM I15066-4230 & 226.673096& -42.515308&  0.547& -0.546& -0.036&  16.8& 0.024& 17.8& 17.5& 17.5& 11.67& 11.19& 10.85&  17.66&  5.99&  d\\
PM I15144-5753 & 228.623672& -57.888840&  0.479& -0.340& -0.337&  18.7& 0.021& 12.0& 10.7& 11.5&  9.47&  8.57&  8.27&  11.34&  1.87& sd\\
PM I15181-6058 & 229.531143& -60.968319&  0.546& -0.293& -0.461&  13.6& 0.029& 14.0& 13.1& 10.9& 10.11&  9.48&  9.27&  14.20&  4.09&  d\\
PM I15217-2713 & 230.442184& -27.217743&  0.594& -0.323& -0.499&  14.9& 0.027& 16.6& 14.5& 13.4& 11.80& 11.34& 11.08&  15.63&  3.83&  d\\
PM I15229-0244 & 230.747040&  -2.748156&  0.611& -0.524& -0.314&  34.2& 0.012& 20.6& 17.6& 15.6& 14.28& 13.80& 13.46&  19.22&  4.94&  d\\
PM I15499-5524 & 237.478027& -55.416222&  0.564& -0.500& -0.261&  10.6& 0.038& 17.1& 15.6& 13.5& 12.63& 12.06& 11.75&  16.41&  3.78& sd\\
PM I15562-2221 & 239.074875& -22.365160&  0.488& -0.269& -0.407&  37.3& 0.011& 15.9& 14.5& 13.4& 12.10& 11.60& 11.37&  15.26&  3.16& sd\\
PM I16006-1654 & 240.171509& -16.908398&  0.812& -0.786&  0.205&  41.9& 0.010& 19.1& 17.5& 17.3& 16.56& 16.27& 16.57&  18.36&  1.80& wd\\
PM I16230-6905 & 245.761459& -69.093605&  1.594& -1.005& -1.237&  17.0& 0.024& 17.5& 15.2& 99.8& 11.38& 10.86& 10.53&  16.44&  5.06&  d\\
PM I16336-3713 & 248.404129& -37.220367&  0.521& -0.517& -0.068&  10.8& 0.037& 16.5& 15.8& 15.8& 15.02& 14.54& 14.34&  16.18&  1.16& wd\\
PM I16389-5555 & 249.743118& -55.918385&  0.642& -0.427& -0.480&  12.0& 0.033& 16.9& 14.8& 14.1& 13.74& 13.23& 13.00&  15.93&  2.19& sd\\
PM I16507-2813 & 252.689224& -28.231855&  0.535& -0.361& -0.395&  13.2& 0.030& 17.4& 13.2& 12.0& 10.69& 10.16&  9.93&  15.47&  4.78&  d\\
PM I16522-2237 & 253.058136& -22.624069&  0.646& -0.262& -0.591&  38.9& 0.010& 16.4& 14.5& 13.9& 12.93& 12.39& 12.17&  15.53&  2.60& sd\\
PM I16527-2913 & 253.180145& -29.218122&  0.484& -0.298& -0.382&  13.2& 0.030& 17.2& 15.8& 14.0& 13.28& 12.70& 12.53&  16.56&  3.28& sd\\
PM I16585-1016 & 254.635452& -10.281461&  0.654& -0.298& -0.582&  37.8& 0.011& 20.4& 18.4& 17.1& 99.99& 99.99& 99.99&  19.48& 99.99&  ?\\
PM I17004-2207 & 255.116562& -22.129393&  0.503& -0.086& -0.496&  38.9& 0.010& 16.2& 13.6& 11.4& 10.22&  9.64&  9.40&  15.00&  4.78&  d\\
PM I17020-3835 & 255.506058& -38.598938&  0.562& -0.204& -0.524&  14.1& 0.028& 18.1& 16.0& 14.5& 14.14& 13.67& 13.54&  17.13&  2.99& sd\\
PM I17032-3104 & 255.809204& -31.081165&  0.561& -0.226& -0.513&  13.2& 0.030& 16.3& 15.1& 13.2& 12.52& 12.00& 11.75&  15.75&  3.23& sd\\
PM I17042-3639 & 256.069702& -36.664280&  0.514& -0.037& -0.513&  11.2& 0.036& 16.2& 14.7& 12.7& 11.53& 11.02& 10.78&  15.51&  3.98&  d\\
PM I17044-2154 & 256.123749& -21.902493&  0.665& -0.501& -0.437&  39.6& 0.010& 17.1& 15.1& 14.0& 13.46& 12.93& 12.75&  16.18&  2.72& sd\\
PM I17046-1257 & 256.170166& -12.953110&  0.674& -0.512& -0.439&  36.9& 0.011& 17.5& 15.3& 14.9& 13.62& 13.09& 12.90&  16.49&  2.87& sd\\
PM I17049-0257 & 256.238464&  -2.951268&  0.493&  0.124& -0.477&  38.1& 0.010& 16.0& 15.9& 15.9& 15.87& 15.75& 15.73&  15.95&  0.08& wd\\
PM I17059-2030 & 256.479004& -20.514038&  0.706& -0.539& -0.456&  37.8& 0.011& 17.8& 15.1& 13.2& 13.42& 12.92& 12.80&  16.56&  3.14& sd\\
PM I17076-2502 & 256.900452& -25.042465&  0.639& -0.204& -0.606&  39.9& 0.010& 16.0& 13.5& 12.0& 11.49& 10.97& 10.77&  14.85&  3.36&  d\\
PM I17082-3436 & 257.061646& -34.606270&  0.861& -0.323& -0.798&  11.2& 0.036& 99.8& 13.1& 10.8& 10.87& 10.36& 10.01&  14.20&  3.33& sd\\
PM I17084-2209W& 257.106445& -22.156849&  0.513& -0.368& -0.358&  39.6& 0.010& 17.7& 15.5& 13.0& 12.92& 12.45& 12.21&  16.80&  3.88&  d\\
PM I17084-2209E& 257.107391& -22.157591&  0.513& -0.368& -0.358&  39.6& 0.010& 17.9& 15.7& 13.1& 13.52& 13.04& 12.79&  16.78&  3.26& sd\\
PM I17088-2105 & 257.215820& -21.087744&  0.890& -0.769& -0.448&  39.6& 0.010& 18.2& 15.2& 13.6& 12.23& 11.77& 11.49&  16.82&  4.59&  d\\
PM I17105-3440 & 257.626434& -34.671501&  0.521&  0.101& -0.511&  11.2& 0.036& 19.2& 17.0& 15.6& 13.55& 13.06& 12.77&  18.19&  4.64&  d\\
PM I17118-1335 & 257.969330& -13.593397&  0.886& -0.488& -0.740&  32.9& 0.012& 15.6& 13.5& 12.5& 10.95& 10.46& 10.18&  14.63&  3.68&  d\\
PM I17137-4535 & 258.431580& -45.599136&  0.466& -0.143& -0.444&  11.9& 0.034& 17.5& 16.1& 15.4& 14.42& 13.92& 13.74&  16.86&  2.44& sd\\
PM I17140-2728 & 258.518677& -27.477440&  0.471& -0.373& -0.287&  13.0& 0.031& 18.5& 17.0& 16.7& 14.93& 13.93& 13.35&  17.81&  2.88& sd\\
PM I17182-1019 & 259.552765& -10.324768&  0.489& -0.269& -0.408&  35.0& 0.011& 18.0& 16.5& 14.3& 12.25& 11.66& 11.33&  17.31&  5.06&  d\\
PM I17189-4131 & 259.727539& -41.527756&  1.160&  0.964& -0.645&  11.8& 0.034& 16.4& 15.4& 12.9& 10.61& 10.00&  9.62&  15.94&  5.33&  d\\
PM I17195-1230 & 259.886536& -12.501122&  0.493&  0.050& -0.490&  35.0& 0.011& 16.2& 14.8& 12.7& 11.27& 10.76& 10.45&  15.56&  4.29&  d\\
PM I17224-1938 & 260.611664& -19.639271&  0.633& -0.322& -0.545&  37.9& 0.011& 17.6& 13.6& 12.1& 11.06& 10.57& 10.23&  15.76&  4.70&  d\\
PM I17238-3541 & 260.962982& -35.686134&  0.569& -0.136& -0.552&  11.2& 0.036& 15.6& 13.5& 11.4& 10.07&  9.49&  9.22&  14.63&  4.56&  d\\
PM I17265-2342 & 261.632538& -23.714851&  0.640& -0.341& -0.542&  42.0& 0.010& 18.4& 16.0& 14.2& 14.34& 13.84& 13.64&  17.30&  2.96& sd\\
PM I17265-5102 & 261.641815& -51.046822&  0.511& -0.245& -0.449&  13.2& 0.030& 15.3& 13.8& 12.2& 11.32& 10.77& 10.55&  14.61&  3.29&  d\\
PM I17273-4851 & 261.833893& -48.866642&  0.610& -0.211& -0.572&  13.2& 0.030& 99.8& 14.1& 12.9& 12.72& 12.25& 12.06&  14.83&  2.11& sd\\
PM I17324-4239 & 263.121307& -42.657875&  0.559& -0.534& -0.165&  11.8& 0.034& 15.8& 14.1& 13.0& 12.23& 11.69& 11.51&  15.02&  2.79& sd\\
PM I17364-3605 & 264.113586& -36.096539&  0.932& -0.390& -0.846&   9.7& 0.041& 14.2& 12.9& 11.4& 10.27&  9.76&  9.52&  13.60&  3.33&  d\\
PM I17386-3427 & 264.654999& -34.458603&  0.491& -0.301& -0.388&   9.7& 0.041& 99.8& 14.1& 13.3& 10.82& 10.24&  9.98&  15.20&  4.38&  d\\
PM I17394-2327 & 264.851440& -23.452009&  0.491& -0.485& -0.074&  11.0& 0.036& 17.7& 13.5& 11.7& 10.62& 10.03&  9.78&  15.77&  5.15&  d\\
PM I17436-2303 & 265.905365& -23.066505&  0.487& -0.080& -0.480&  11.0& 0.036& 17.1& 14.1& 12.4& 11.93& 11.25& 11.03&  15.72&  3.79&  d\\
PM I17444-2044 & 266.121948& -20.747021&  0.736& -0.240& -0.696&  41.1& 0.010& 16.8& 13.9& 12.3& 11.31& 10.77& 10.60&  15.47&  4.16&  d\\
PM I17468-0346 & 266.711670&  -3.769525&  0.535& -0.425& -0.325&  35.0& 0.011& 18.3& 16.8& 15.5& 14.49& 13.99& 13.89&  17.61&  3.12& sd\\
PM I17476-5436 & 266.903381& -54.608665&  0.462& -0.392& -0.244&   3.0& 0.133& 15.7& 15.3& 14.4& 14.46& 14.22& 14.29&  15.52&  1.06& wd\\
PM I17480-5043 & 267.012421& -50.728275&  0.577& -0.200& -0.541&  13.2& 0.030& 17.3& 15.3& 13.5& 12.55& 12.02& 11.81&  16.38&  3.83&  d\\
PM I17489-4002 & 267.232697& -40.039345&  0.523& -0.289& -0.436&  12.3& 0.033& 15.1& 13.8& 12.5& 12.35& 11.83& 11.64&  14.50&  2.15& sd\\
PM I17509-2734 & 267.727722& -27.576120&  0.533& -0.403& -0.349&  11.0& 0.036& 15.5& 13.3& 13.4& 11.73& 11.21& 10.95&  14.49&  2.76& sd\\
PM I17521-3916 & 268.046082& -39.281761&  0.491& -0.130& -0.473&  12.3& 0.033& 15.8& 14.8& 13.1& 11.83& 11.24& 10.99&  15.34&  3.51&  d\\
PM I17525-0248 & 268.140717&  -2.811024&  0.530& -0.063& -0.526&  30.0& 0.013& 16.6& 14.8& 13.2& 12.10& 11.58& 11.37&  15.77&  3.67&  d\\
PM I17540-3440 & 268.512390& -34.671764&  0.639& -0.465& -0.438&  10.0& 0.040& 15.2& 12.9& 11.6&  9.43&  8.88&  8.55&  14.14&  4.71&  d\\
PM I17595-1755 & 269.885742& -17.926767&  1.143& -0.410& -1.067&  37.0& 0.011& 15.5& 13.9& 13.0& 11.44& 10.94& 10.69&  14.76&  3.32& sd\\
PM I18018-2706 & 270.469421& -27.115088&  0.557& -0.243& -0.501&  41.1& 0.010& 16.2& 13.2& 12.8& 10.51&  9.90&  9.60&  14.82&  4.31&  d\\
PM I18044-1005 & 271.117157& -10.091301&  0.483& -0.283& -0.392&  34.2& 0.012& 17.0& 14.7& 99.8& 11.41& 10.84& 10.53&  15.94&  4.53&  d\\
PM I18063-1913 & 271.575317& -19.224737&  0.851& -0.557& -0.643&  37.1& 0.011& 17.2& 14.4& 14.4& 12.95& 12.38& 12.15&  15.91&  2.96& sd\\
PM I18087-2632 & 272.198090& -26.544395&  0.559& -0.322& -0.457&  12.2& 0.033& 14.7& 12.9& 99.8& 10.30&  9.73&  9.49&  13.87&  3.57&  d\\
PM I18088-7042 & 272.223938& -70.707069&  0.523&  0.213& -0.478&  14.9& 0.027& 18.4& 16.3& 15.8& 15.74& 15.37& 15.41&  17.43&  1.69& wd\\
PM I18094-4104 & 272.352509& -41.076588&  0.481&  0.237& -0.419&  12.0& 0.033& 16.0& 14.6& 12.9& 11.43& 10.90& 10.61&  15.36&  3.93&  d\\
PM I18095-3012 & 272.376709& -30.205776&  0.486& -0.374& -0.310&  11.8& 0.034& 13.7& 12.6& 11.0& 10.46&  9.91&  9.69&  13.19&  2.73&  d\\
PM I18106-0650 & 272.667755&  -6.848365&  0.484& -0.136& -0.464&  34.1& 0.012& 16.0& 14.0& 12.9& 12.07& 11.51& 11.39&  15.08&  3.01& sd\\
PM I18118-1419 & 272.950073& -14.325014&  0.455& -0.445& -0.095&  34.1& 0.012& 19.6& 17.3& 17.0& 15.21& 13.87& 13.38&  18.54&  3.33& sd\\
PM I18177-1302 & 274.440796& -13.045858&  0.483& -0.301& -0.378&  32.2& 0.012& 18.6& 16.0& 14.8& 13.35& 12.67& 11.83&  17.40&  4.05&  d\\
PM I18200-3828 & 275.012390& -38.472858&  0.747& -0.335& -0.668&  12.0& 0.033& 15.5& 13.9& 12.3& 10.92& 10.39& 10.13&  14.76&  3.84&  d\\
PM I18244-2521 & 276.123779& -25.352663&  0.642& -0.318& -0.558&  12.2& 0.033& 14.9& 14.7& 11.4& 10.66& 10.01&  9.74&  14.81&  4.15&  d\\
PM I18265-1923 & 276.631653& -19.391891&  0.552& -0.282& -0.474&  39.9& 0.010& 14.4& 12.9& 11.0&  9.77&  9.17&  8.92&  13.71&  3.94&  d\\
PM I18292-3059S& 277.319672& -30.998821&  0.484&  0.457& -0.158&  11.8& 0.034& 99.8& 12.5& 99.8& 10.67& 10.16&  9.94&  13.50&  2.83&  d\\
PM I18292-3059N& 277.320496& -30.996946&  0.484&  0.457& -0.158&  11.8& 0.034& 99.8& 12.4& 99.8& 10.25&  9.77&  9.49&  13.50&  3.25&  d\\
PM I18310-2939 & 277.770935& -29.661570&  0.750& -0.051& -0.748&  11.8& 0.034& 14.6& 13.3& 11.3& 10.34&  9.86&  9.59&  14.00&  3.66&  d\\
PM I18429-1639 & 280.734833& -16.656038&  0.473& -0.369& -0.296&  35.6& 0.011& 16.4& 16.1& 12.7& 12.34& 11.87& 11.57&  16.26&  3.92&  d\\
PM I18456-2337 & 281.405945& -23.630495&  0.646&  0.214& -0.609&  11.9& 0.034& 17.1& 14.5& 13.5& 11.51& 10.97& 10.73&  15.90&  4.39&  d\\
PM I18530-1533 & 283.269012& -15.551133&  0.506& -0.091& -0.498&  33.0& 0.012& 15.8& 14.2& 13.2& 11.19& 10.70& 10.48&  15.06&  3.87&  d\\
PM I18533-2556 & 283.332947& -25.945759&  0.493& -0.302& -0.390&  41.8& 0.010& 20.3& 18.0& 16.6& 15.29& 14.73& 14.53&  19.24&  3.95& sd\\
PM I18536-2305 & 283.417114& -23.090977&  0.455&  0.361& -0.277&  11.9& 0.034& 18.5& 15.9& 16.6& 14.87& 14.42& 14.07&  17.30&  2.43& sd\\
PM I19068-2414 & 286.706268& -24.245245&  0.655&  0.097& -0.648&   3.8& 0.105& 17.4& 15.7& 14.2& 12.64& 12.13& 11.91&  16.62&  3.98& sd\\
PM I19072-1931 & 286.819061& -19.531000&  0.562& -0.277& -0.489&  37.0& 0.011& 18.0& 16.4& 15.1& 14.73& 14.24& 14.10&  17.26&  2.53& sd\\
PM I19131-1151 & 288.280762& -11.862134&  0.598& -0.121& -0.586&  36.7& 0.011& 16.0& 13.8& 12.4& 11.08& 10.49& 10.33&  14.99&  3.91&  d\\
PM I19163-2322 & 289.096924& -23.375906&  0.572&  0.251& -0.514&  37.0& 0.011& 16.8& 13.9& 11.2& 10.20&  9.63&  9.33&  15.47&  5.27&  d\\
PM I19316-0658 & 292.911163&  -6.973796&  0.544& -0.143& -0.525&  34.7& 0.012& 15.2& 12.8& 10.6&  9.48&  8.77&  8.58&  14.10&  4.62&  d\\
PM I19334-1434 & 293.372406& -14.580217&  0.886& -0.571& -0.677&  34.0& 0.012& 14.0& 12.3& 11.4& 10.63& 10.07&  9.85&  13.22&  2.59& sd\\
PM I19350-0725 & 293.756409&  -7.424497&  0.718&  0.536& -0.477&  33.7& 0.012& 17.1& 14.9& 13.6& 12.03& 11.53& 11.25&  16.09&  4.06&  d\\
PM I19394-1016 & 294.851349& -10.280749&  0.466& -0.252& -0.392&  35.9& 0.011& 18.5& 16.5& 15.1& 13.88& 13.38& 13.17&  17.58&  3.70& sd\\
PM I19418-0208 & 295.470551&  -2.146091&  1.093& -0.282& -1.056&  35.1& 0.011& 18.0& 16.9& 16.6& 14.62& 14.14& 13.99&  17.49&  2.87& sd\\
PM I19427-1723 & 295.680054& -17.390104&  0.904&  0.028& -0.904&  35.1& 0.011& 14.1& 13.0& 10.6&  9.83&  9.26&  8.98&  13.59&  3.76&  d\\
PM I19428-0304 & 295.700165&  -3.075746&  0.868& -0.549& -0.672&  31.9& 0.013& 18.6& 16.4& 14.7& 13.98& 13.45& 13.25&  17.59&  3.61& sd\\
PM I19484-5101 & 297.117523& -51.025124&  0.690&  0.190& -0.663&  15.2& 0.026& 99.8& 16.1& 13.9& 12.46& 11.98& 11.68&  17.20&  4.74&  d\\
PM I19578-2513 & 299.468140& -25.225792&  0.599& -0.092& -0.591&  15.1& 0.026& 18.8& 17.9& 16.9& 15.24& 14.99& 14.78&  18.39&  3.15& sd\\
PM I20079-4205 & 301.980164& -42.087429&  0.771&  0.044& -0.770&  17.1& 0.023& 15.5& 13.6& 10.9&  9.52&  8.96&  8.61&  14.63&  5.11&  d\\
PM I20138-0216 & 303.468933&  -2.272357&  0.498& -0.133& -0.480&  37.0& 0.011& 19.3& 16.3& 14.6& 13.00& 12.52& 12.23&  17.92&  4.92&  d\\
PM I20205-4202 & 305.128723& -42.048962&  0.473&  0.135& -0.453&  14.6& 0.027& 16.7& 16.3& 16.5& 16.34& 15.98& 15.86&  16.52&  0.18& wd\\
PM I20271-4421 & 306.779480& -44.358181&  0.550&  0.178& -0.520&  13.9& 0.029& 15.0& 13.3& 11.5& 10.08&  9.48&  9.16&  14.40&  4.32&  d\\
PM I20278-4301 & 306.956390& -43.020847&  0.823&  0.199& -0.799&  13.9& 0.029& 17.9& 16.7& 16.5& 15.85& 15.61& 15.61&  17.35&  1.50& wd\\
PM I20289-5004 & 307.243835& -50.071632&  0.484&  0.279& -0.396&  14.9& 0.027& 14.0& 12.9& 11.8& 11.04& 10.46& 10.26&  13.92&  2.88&  d\\
PM I20335-3735 & 308.389313& -37.595295&  0.528&  0.513& -0.125&  14.6& 0.027& 19.3& 17.1& 14.6& 12.74& 12.21& 11.86&  18.29&  5.55&  d\\
PM I20347-3630 & 308.678497& -36.515709&  0.693&  0.189& -0.667&  15.2& 0.026& 99.8& 15.3& 99.8& 12.27& 11.82& 11.54&  16.40&  4.13&  d\\
PM I20348-6347 & 308.720459& -63.795998&  0.452&  0.238& -0.384&  15.0& 0.027& 14.5& 12.1& 10.6&  9.37&  8.76&  8.51&  13.40&  4.03&  d\\
PM I20457-1411 & 311.429321& -14.192146&  0.814&  0.157& -0.799&  31.8& 0.013& 20.5& 17.5& 99.8& 15.10& 14.57& 14.57&  19.12&  4.02& sd\\
PM I20457-3013 & 311.448883& -30.223770&  0.512&  0.512& -0.009&  16.2& 0.025& 17.8& 15.9& 13.5& 11.77& 11.22& 10.91&  16.93&  5.16&  d\\
PM I20502-3424 & 312.567383& -34.411858&  0.533&  0.360& -0.393&  14.9& 0.027& 14.8& 12.8& 10.4&  8.82&  8.27&  8.00&  13.88&  5.06&  d\\
PM I20589-3457 & 314.743744& -34.952717&  0.687&  0.339& -0.597&  14.9& 0.027& 15.1& 13.8& 13.2& 11.96& 11.52& 11.38&  14.50&  2.54& sd\\
PM I21023-6046 & 315.592163& -60.771648&  0.560&  0.168& -0.534&  15.1& 0.026& 19.6& 17.6& 17.0& 15.63& 15.20& 14.83&  18.68&  3.05& sd\\
PM I21163-3303 & 319.076721& -33.063358&  0.768&  0.712& -0.289&  14.9& 0.027& 99.8& 15.9& 13.9& 12.39& 11.86& 11.58&  17.00&  4.61&  d\\
PM I21340-1413 & 323.510071& -14.228679&  0.648& -0.314& -0.567&  33.2& 0.012& 20.6& 18.3& 15.6& 13.99& 13.43& 13.11&  19.54&  5.55&  d\\
PM I21349-0500 & 323.736084&  -5.015940&  0.466& -0.306& -0.351&  31.1& 0.013& 18.0& 17.5& 16.8& 16.70& 16.03& 16.28&  17.77&  1.07& wd\\
PM I21356-3448 & 323.910217& -34.812588&  0.639&  0.626& -0.127&  15.1& 0.026& 15.8& 13.9& 12.2& 11.20& 10.67& 10.44&  14.93&  3.73&  d\\
PM I21375-1014 & 324.379944& -10.240491&  0.625&  0.181& -0.598&  34.2& 0.012& 16.1& 14.1& 99.8& 10.26&  9.73&  9.43&  15.18&  4.92&  d\\
PM I22033-2359 & 330.843201& -23.995905&  0.513&  0.503& -0.102&  14.1& 0.028& 18.5& 16.9& 14.4& 12.53& 12.00& 11.66&  17.76&  5.23&  d\\
PM I23035-2809 & 345.884521& -28.157280&  0.464&  0.463&  0.016&  14.1& 0.028& 20.3& 17.0& 15.6& 13.61& 13.06& 12.80&  18.78&  5.17&  d\\
PM I23187-3135 & 349.684784& -31.583717&  0.511&  0.456& -0.229&  19.0& 0.021& 19.8& 16.7& 99.8& 99.99& 99.99& 99.99&  18.37& 99.99&  ?\\
PM I23440-8246 & 356.014435& -82.782303&  0.694&  0.694& -0.022&  18.8& 0.021& 17.2& 16.6& 16.6& 15.97& 15.73& 15.53&  16.92&  0.95& wd\\
\enddata                              
\tablenotetext{a}{Temporal baseline between the first and second DSS epoch.}
\tablenotetext{b}{Estimated proper motion error.}
\tablenotetext{c}{Photographic $B_J$ (IIIaJ), $R_F$ (IIIaF) and $I_N$
(IVN) magnitudes from the USNO-B1.0 catalog.}
\tablenotetext{d}{Infrared $JHK_s$ magnitudes from the
2MASS All-Sky Point Source Catalog.}
\end{deluxetable}   
\clearpage
\end{landscape}

\begin{deluxetable}{lrrrrr}
\tabletypesize{\scriptsize}
\tablecolumns{6} 
\tablecaption{Recovery rate by SUPERBLINK of known high proper motion 
  stars from various catalogs\tablenotemark{1}}
 \tablehead{Source & southern sky \# \tablenotemark{2}&  recovered & missed & recovery rate}
\startdata 
Giclas catalog     &  247 &  235 & 12 & 95.1\% \\
Luyten catalogs    & 1477 & 1395 & 82 & 94.4\% \\
WT/WC survey       &  152 &  139 & 13 & 91.4\% \\
Cal\`an-ESO survey   &   46 &   42 &  4 & 91.3\% \\
APMPM survey       &  143 &  118 & 25 & 82.5\% \\
LEHPM survey       &  331 &  286 & 45 & 86.4\% \\
SuperCOSMOS-RECONS &  196 &  174 & 22 & 88.8\% \\
SIPS survey        &   68 &   20 & 48 & 29.4\% \\
\enddata
\tablenotetext{1}{Stars in the southern sky, with proper motions
  $0.45\arcsec$ yr$^{-1}<\mu<2.0\arcsec$ yr$^{-1}$, and visual
  magnitude $V>10$.}
\tablenotetext{2}{Total number of objects in the Southern sky. Surveys
  are not mutually exclusive; stars can be tabulated in several of
  these lists.}
\end{deluxetable}

\appendix

\section{Finder charts}

Finder charts of HPM stars are generated as a
by-product of the SUPERBLINK software. Figure 6 presents finder charts
for all the new, HPM stars presented in this paper and
listed in Table 1. Each chart consists in a pair of images showing the
local field at two different epochs. The name of the star is indicated
in the center just below the chart, and corresponds to the name given
in Table 1. To the left is the POSS-I field, with the epoch of the
field noted in the lower left corner. To the right is the modified
POSS-II field which has been shifted, rotated, and degraded in such a
way that it matches the quality and aspect of the POSS-I image. The
epoch of the POSS-II field is noted on the lower right corner. HPM
stars are identified with circles centered on their
positions at the epoch on the plate.

The charts are oriented in the local X-Y coordinate system of the
POSS-I image; the POSS-II image has been remapped on the POSS-I
grid. The specific orientation of each field is specified by a
north-east (N-E) compass, drawn on the first epoch panel. Each segment
of the compass is 1 arcmin in length, which provides a sense of
scale. Most of the charts are $4.25\arcmin\times4.25\arcmin$ on the
side, but a few are $2.5\arcmin\times2.5\arcmin$ on the side. These
are easily distinguished by the size of the N-E compass. 

Sometimes a part of the field is missing: this is an artifact
of the code. SUPERBLINK works on $17\arcmin\times17\arcmin$ DSS
subfields. If an HPM star is identified near the edge of
that subfield, the output chart appears truncated.



\begin{figure}
\epsscale{1.0}
\plotone{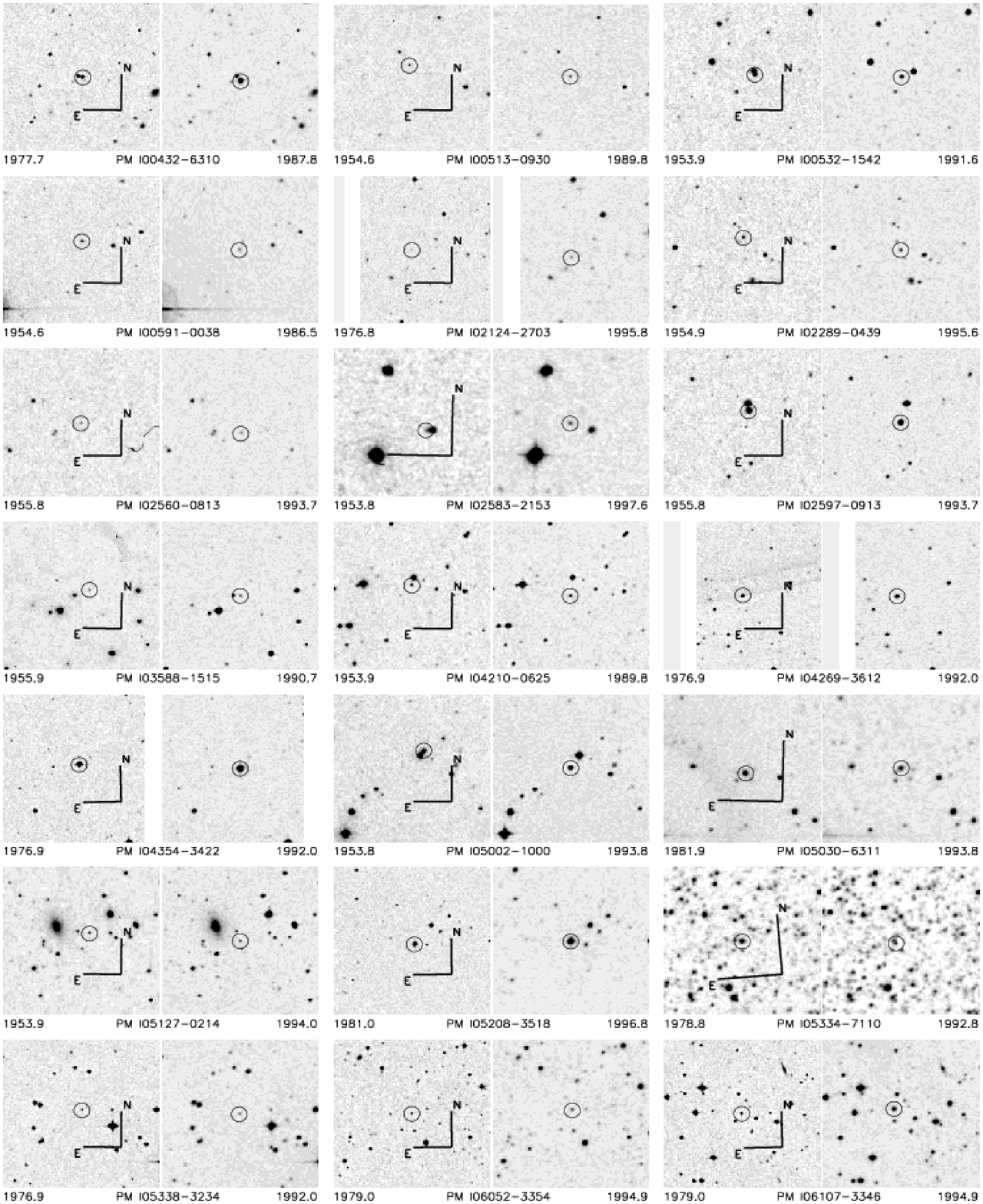}
\caption{Finder charts for the new HPM stars
discovered with SUPERBLINK, as listed in Table 1.}
\end{figure}

\begin{figure}
\epsscale{1.0}
\plotone{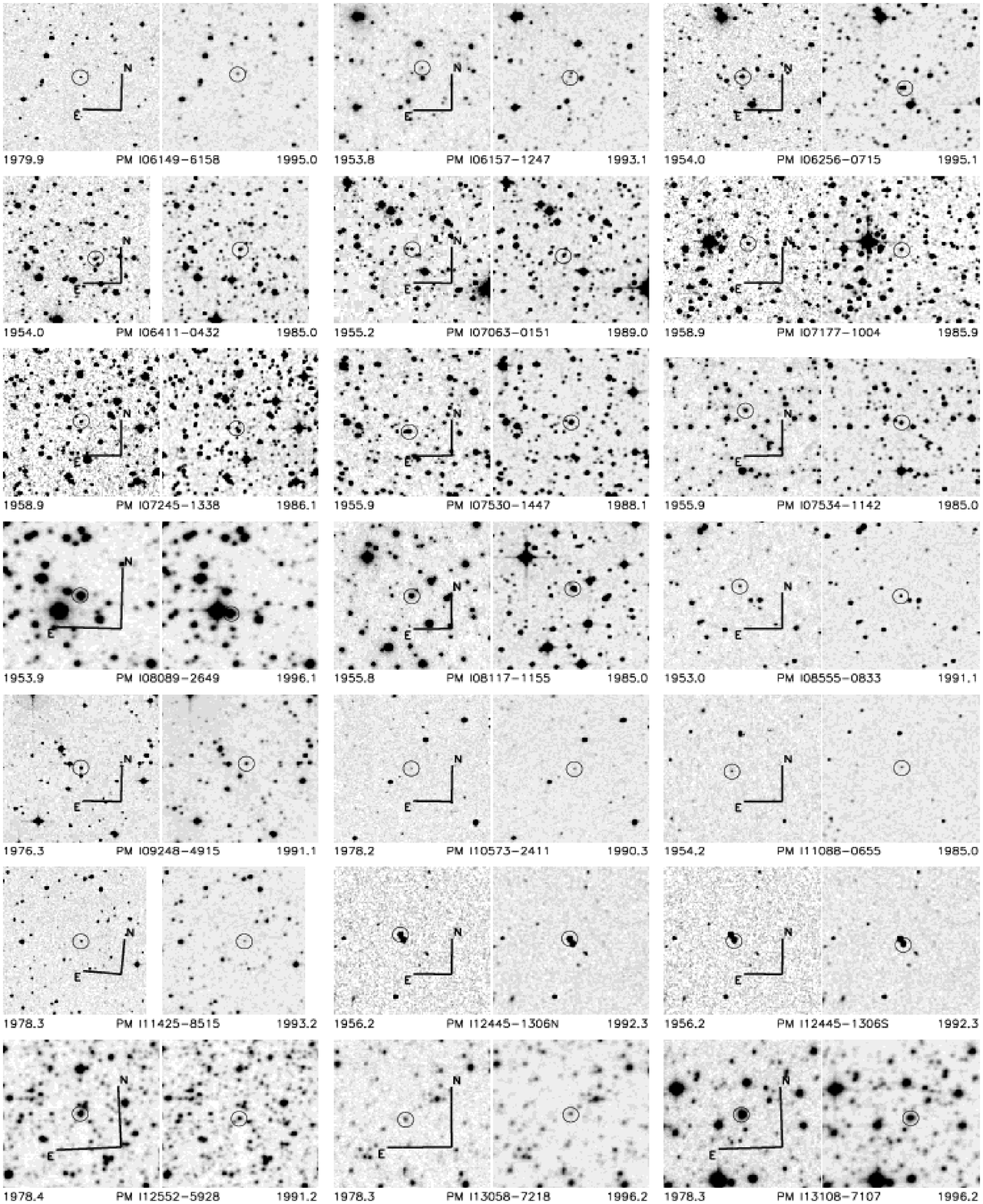}
\caption{Finder charts for the new HPM stars
discovered with SUPERBLINK (continued).}
\end{figure}

\begin{figure}
\epsscale{1.0}
\plotone{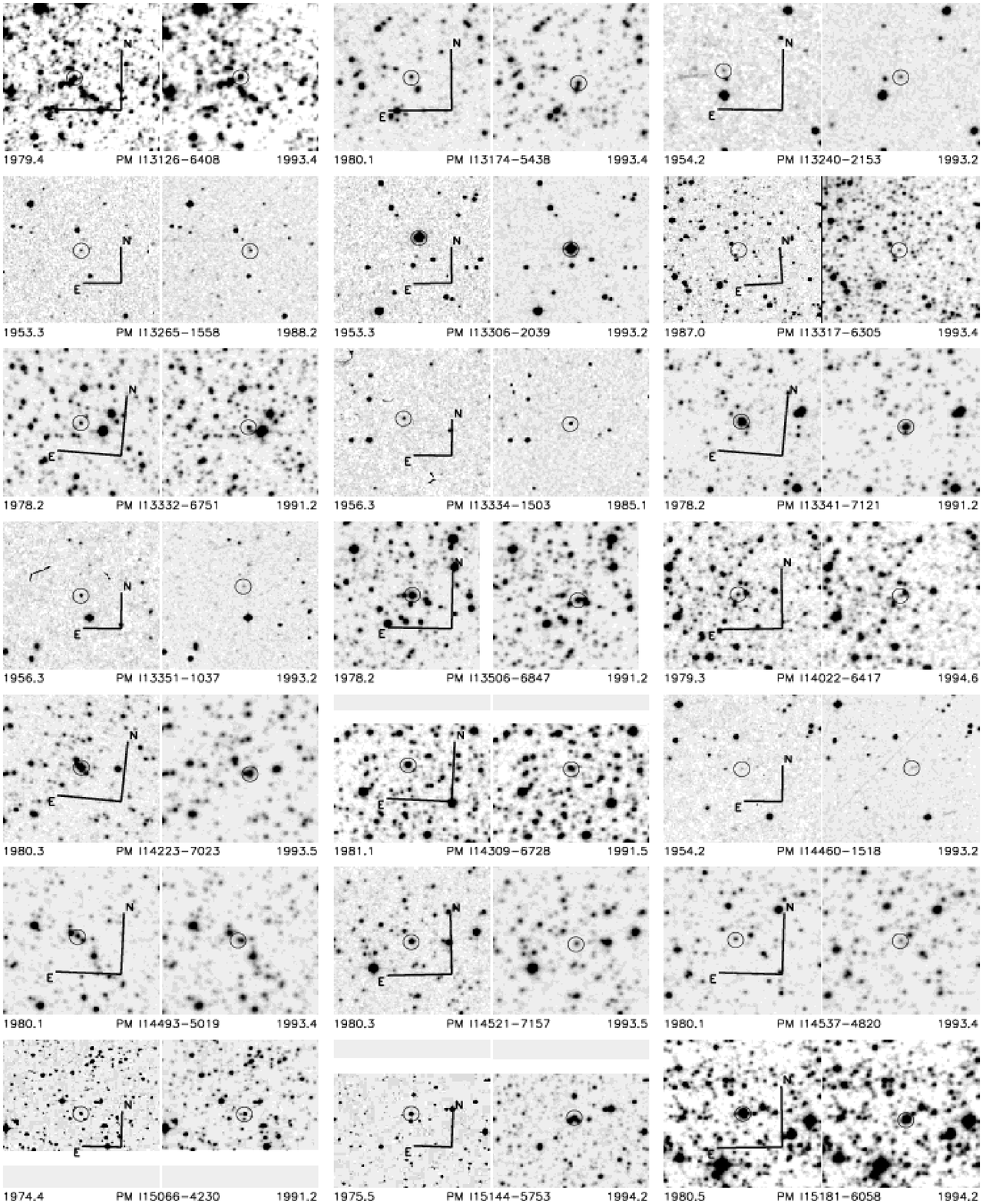}
\caption{Finder charts for the new HPM stars
discovered with SUPERBLINK (continued).}
\end{figure}

\begin{figure}
\epsscale{1.0}
\plotone{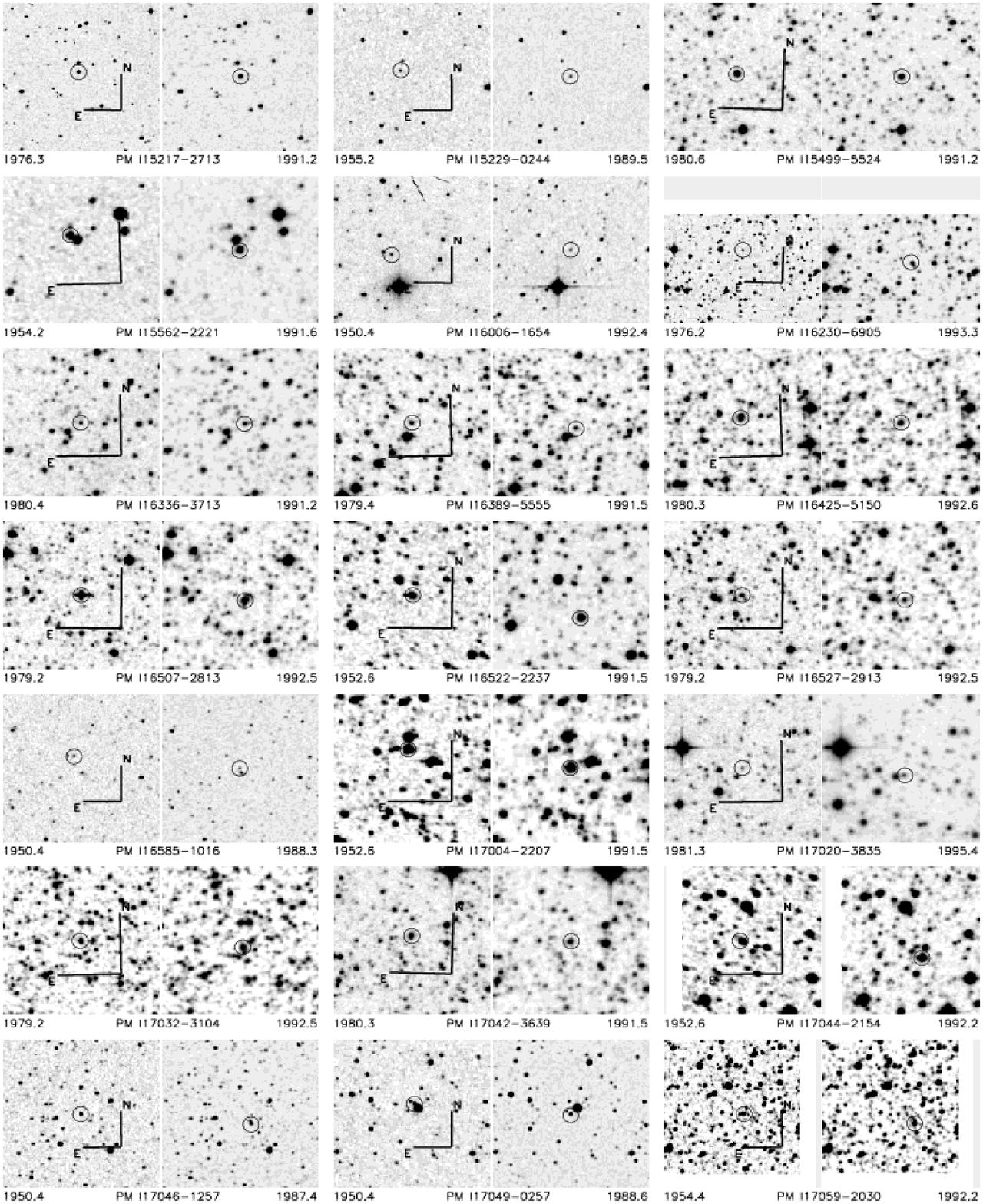}
\caption{Finder charts for the new HPM stars
discovered with SUPERBLINK (continued).}
\end{figure}

\begin{figure}
\epsscale{1.0}
\plotone{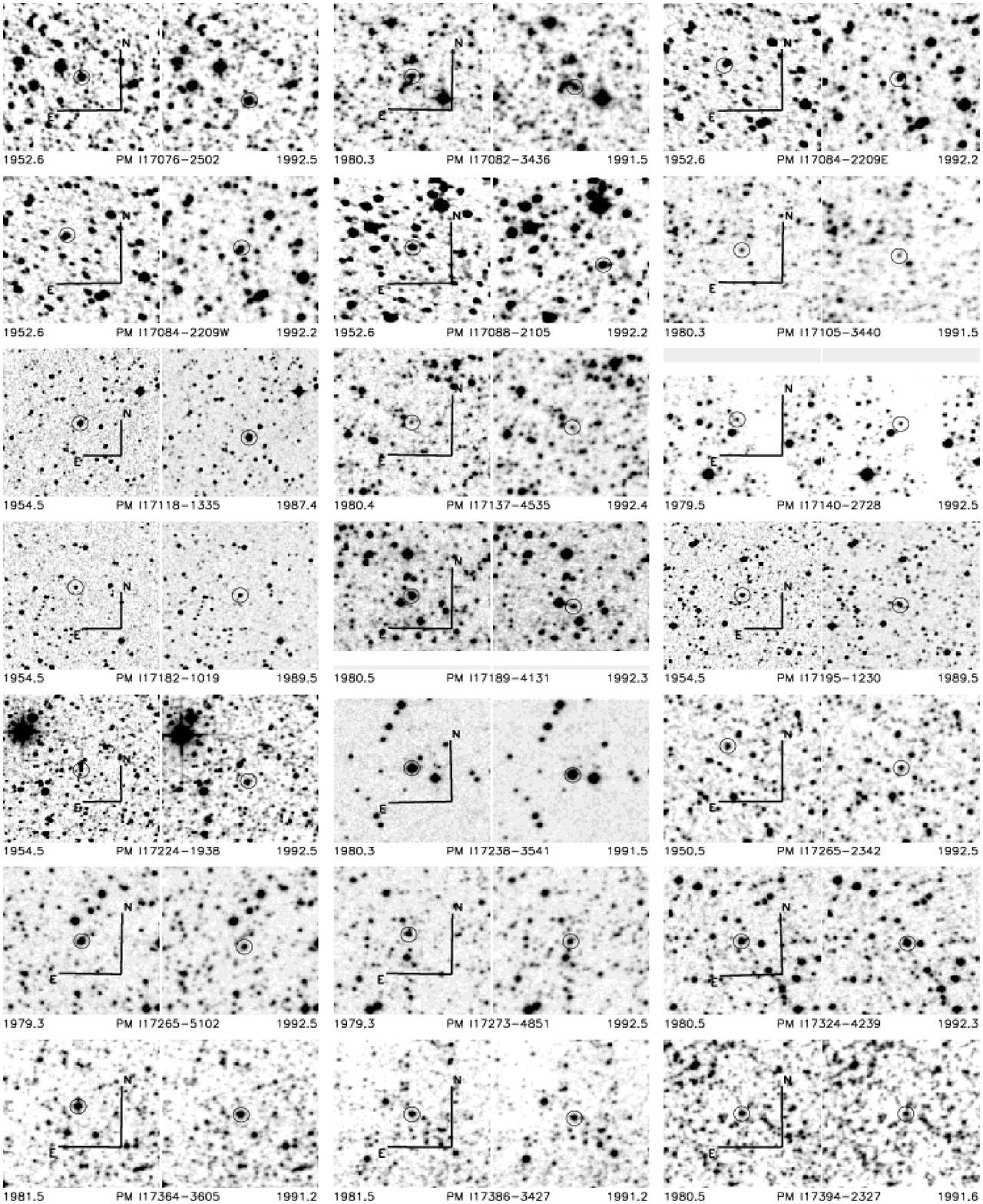}
\caption{Finder charts for the new HPM stars
discovered with SUPERBLINK (continued).}
\end{figure}

\begin{figure}
\epsscale{1.0}
\plotone{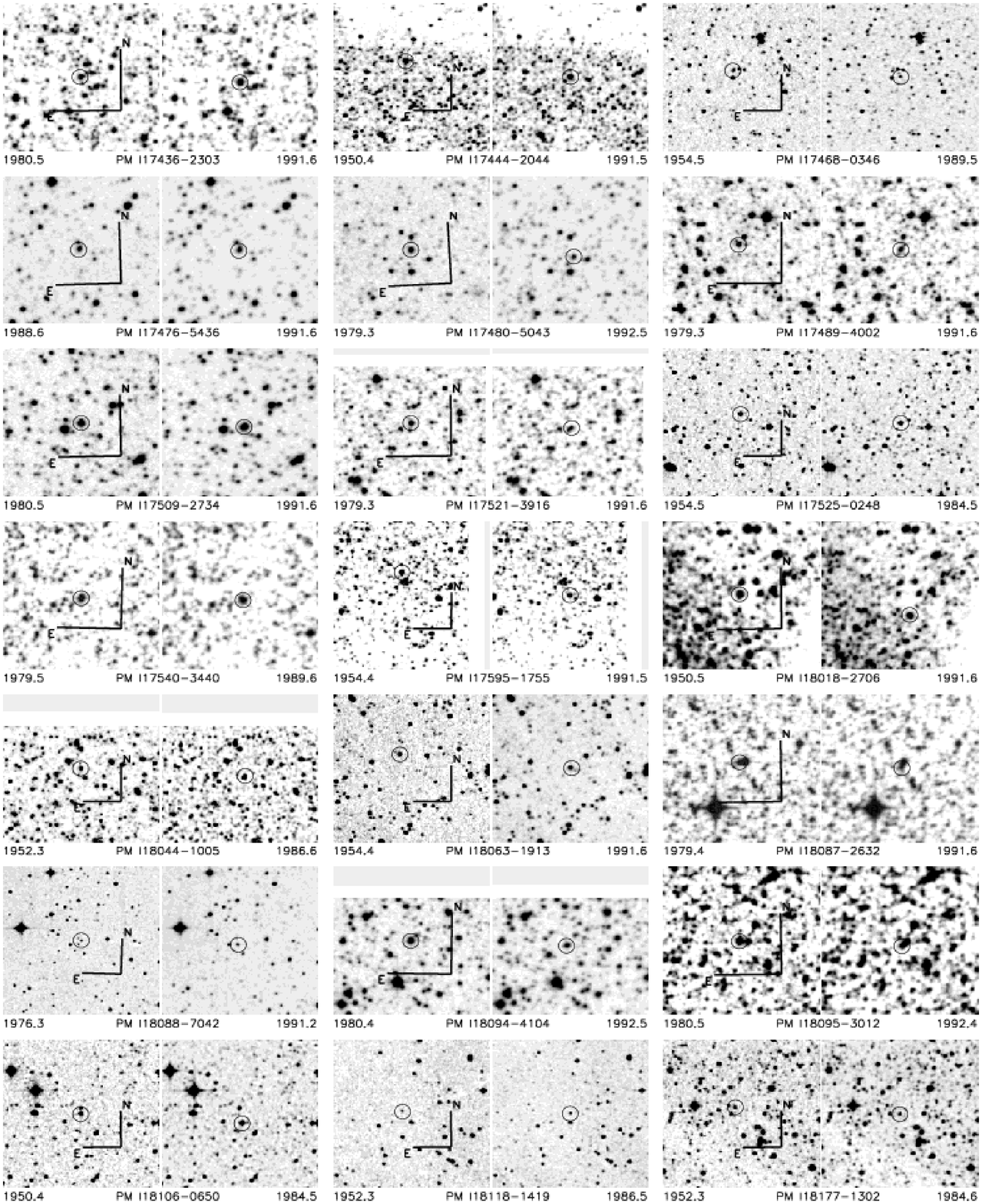}
\caption{Finder charts for the new HPM stars
discovered with SUPERBLINK (continued).}
\end{figure}

\begin{figure}
\epsscale{1.0}
\plotone{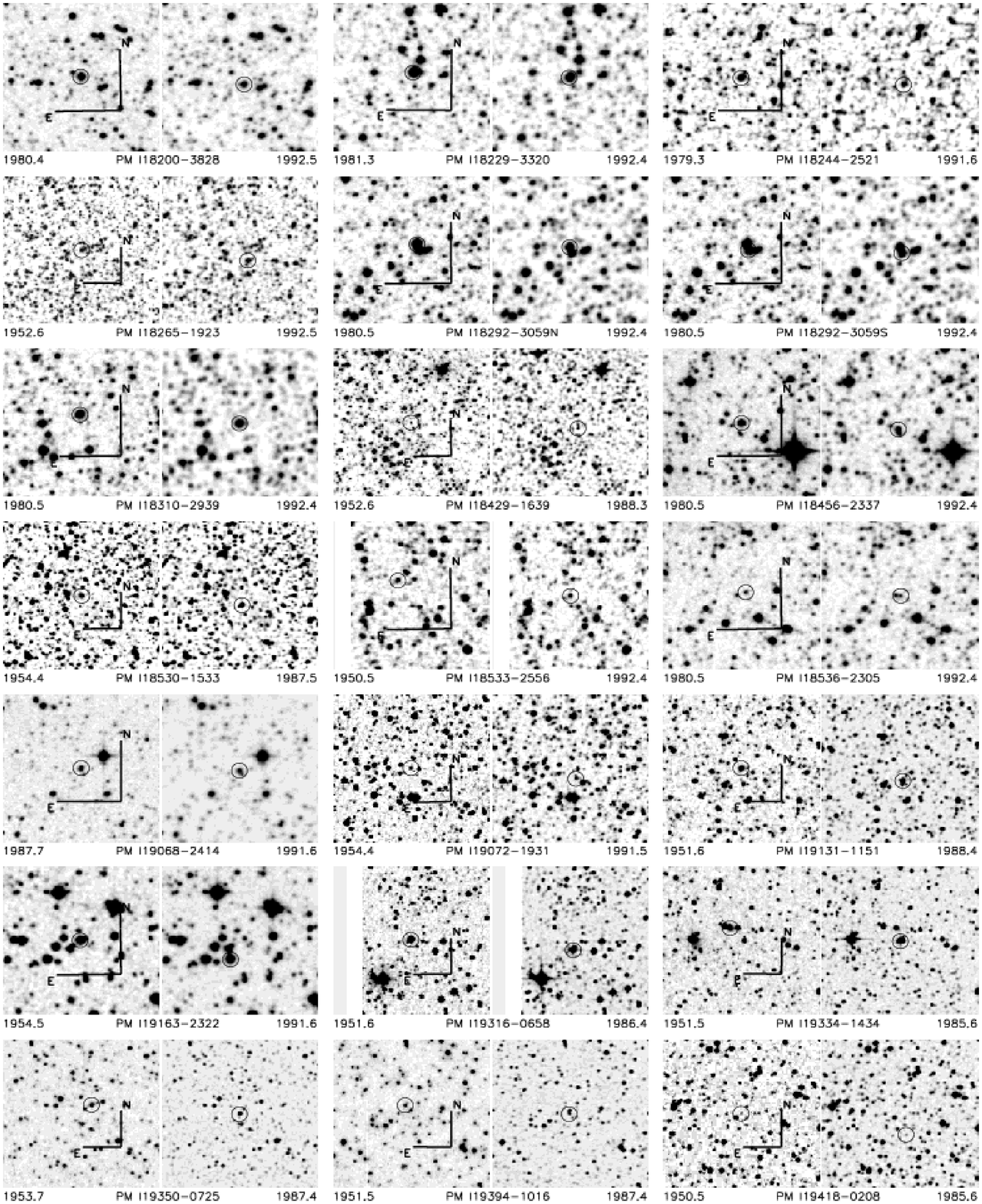}
\caption{Finder charts for the new HPM stars
discovered with SUPERBLINK (continued).}
\end{figure}

\begin{figure}
\epsscale{1.0}
\plotone{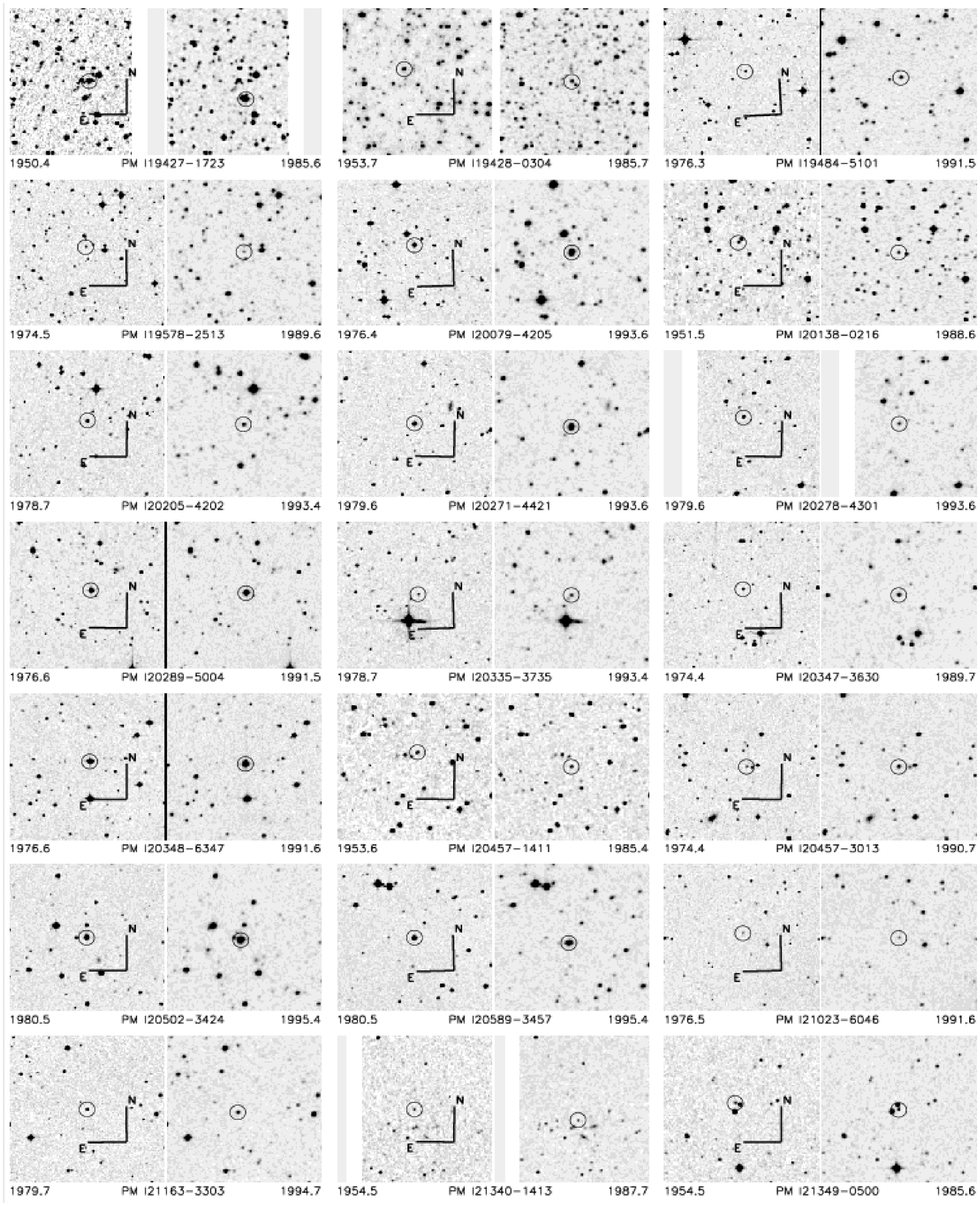}
\caption{Finder charts for the new HPM stars
discovered with SUPERBLINK (continued).}
\end{figure}

\begin{figure}
\epsscale{1.0}
\plotone{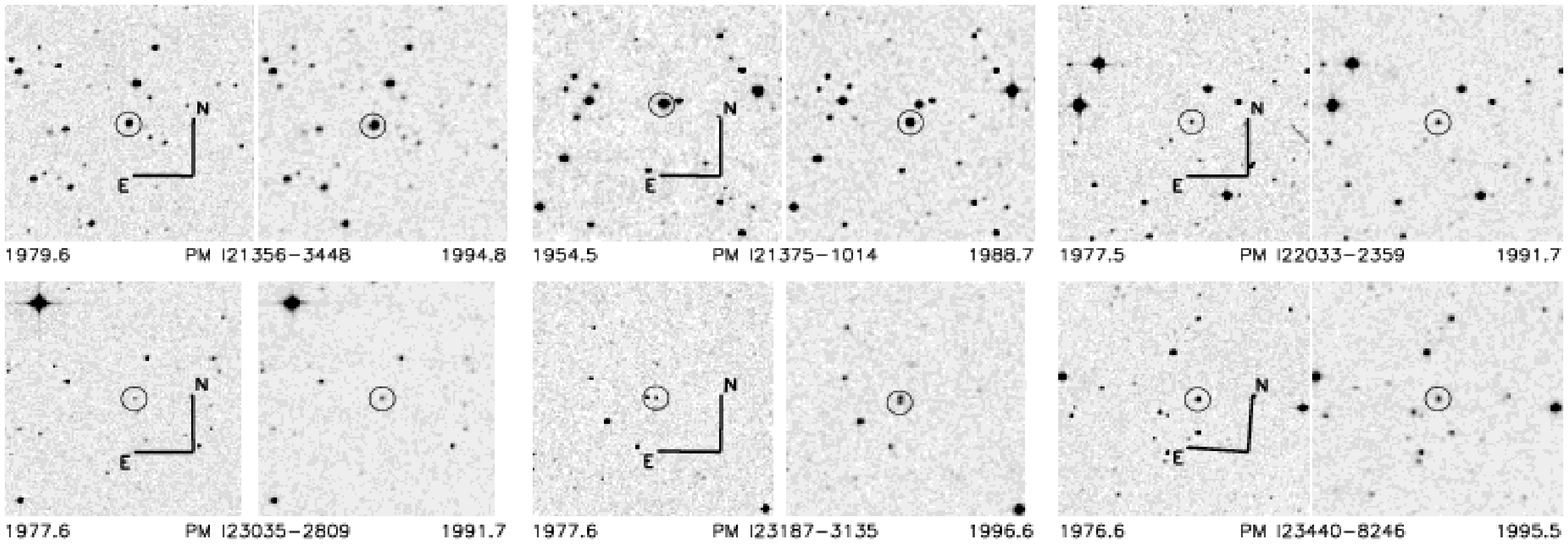}
\caption{Finder charts for the new HPM stars
discovered with SUPERBLINK (continued).}
\end{figure}

\end{document}